\documentclass[prc,letterpaper,twocolumn,showpacs,showkeys,lengthcheck,tightenlines,
               nofootinbib,preprintnumbers,superscriptaddress]{revtex4-1}

\usepackage[utf8]{inputenc}

\usepackage{titlesec}
\usepackage{dcolumn}
\usepackage{bm}
\usepackage{amsmath,amssymb,amsfonts}

\usepackage{graphicx}
\usepackage{subfigure}
\usepackage{color}
\usepackage{bbold}

\usepackage{slashed}
\usepackage[svgnames]{xcolor}
\usepackage[english]{babel}
\usepackage{blindtext}
\usepackage{microtype}
\usepackage{tikz}

\usepackage[colorlinks=true,urlcolor=blue,citecolor=blue]{hyperref}
\usepackage{ifpdf}
\usepackage{float}

\graphicspath{{.}{figures/}}

\titlespacing{\section}{5pt}{12pt plus 4pt minus 2pt}{8pt plus 2pt minus 2pt}
\titlespacing{\subsection}{0pt}{12pt plus 4pt minus 2pt}{8pt plus 2pt minus 2pt}

\begin{document}

\title{$\rho$ meson form factors and parton distribution functions in impact parameter space}

\author{Jin-Li Zhang}
\email[]{jlzhang@njit.edu.cn}
\affiliation{Department of Mathematics and Physics, Nanjing Institute of Technology, Nanjing 211167, China }

\begin{abstract}
In this paper, we investigate the form factors and impact parameter space parton distribution functions of the $\rho$ meson derived from the generalized parton distributions within the framework of the Nambu–Jona-Lasinio model, employing a proper time regularization scheme. We compare the charge $G_C$, magnetic $G_M$, and quadrupole $G_Q$ form factors with lattice data. The dressed form factors, $G_C^D$ and $G_M^D$, exhibit good agreement with lattice results; however, $G_Q^D$ is found to be harder than what is observed in lattice calculations. The Rosenbluth cross section for elastic electron scattering on a spin-one particle can be expressed through the structure functions $A(Q^2)$ and $B(Q^2)$. Additionally, the tensor polarization $T_{20}(Q^2,\theta)$ can also be formulated in terms of these form factors. We analyze the structure functions $A(Q^2)$, $B(Q^2)$ and tensor polarization function $T_{20}(Q^2,\theta)$; our findings quantitatively align with predicted values across various limits. In impact parameter space, we examine parton distribution functions along with their dependence on longitudinal momentum fraction $x$ and impact parameter $\bm{b}_{\perp}$. The width distributions in impact parameter space reveal that the range of the charge distribution \( q_C(x,\bm{b}_{\perp}^2) \) is the most extensive. In contrast, the transverse magnetic radius falls within a moderate range, while the quadrupole distribution \( q_Q(x,\bm{b}_{\perp}^2) \) demonstrates the narrowest extent.
\end{abstract}
\maketitle
\section{Introduction }
The investigation into the inner structure of matter and the fundamental laws governing interactions has consistently been at the forefront of natural science research. This pursuit not only enables humanity to comprehend the underlying principles of nature but also fosters significant advancements in various technologies. Multidimensional imaging of hadrons has generated significant interest over the past several decades. It is widely acknowledged that generalized parton distributions (GPDs)~\cite{Mueller:1998fv,Ji:1996ek,Radyushkin:1997ki,Vanderhaeghen:1999xj,Belitsky:2001ns,Burkardt:2002hr,Diehl:2003ny,Sun:2017gtz,Zhang:2020ecj,Zhang:2021mtn,Zhang:2021shm,Zhang:2021tnr,Zhang:2021uak,Zhang:2022zim,Zhang:2023xfe,Moffat:2023svr,Goloskokov:2022mdn,Tan:2023kbl,Lin:2023ezw,Zhang:2024adr} and transverse momentum dependent parton distribution functions (TMDs)~\cite{Pasquini:2008ax,Musch:2010ka,Lorce:2011dv,Angeles-Martinez:2015sea,Echevarria:2016scs,Bacchetta:2017gcc,Ninomiya:2017ggn,Zhang:2024plq,Cheng:2024gyv,Kou:2023ady,Zhu:2022bja,Zeng:2022lbo,Deng:2022gzi,LatticeParton:2023xdl,Cheng:2024gyv} serve as a powerful tool for elucidating the hadronic structure of a system. This efficacy arises from the fact that GPDs inherently encapsulate information pertaining to both form factors (FFs)~\cite{Lepage:1979zb,JeffersonLabHallA:2001qqe,Hohler:1976ax,Gasser:1984ux,Brodsky:1992px,CLEO:1997fho,Ball:2004rg,Bednar:2018mtf,Alexandrou:2021ztx,Saha:2024jkf,Kumar:2024eui,Sultan:2024hep,Higuera-Angulo:2024oui,Gurjar:2024wpq,Dbeyssi:2011ep,Hernandez-Pinto:2024kwg} and parton distribution functions (PDFs)~\cite{jaffe1983parton,Duke:1983gd,soper1997parton,giele2001parton,Martin:2009bu,Botje:2011sn,gao2014meta,Radyushkin:2017cyf,nagy2020evolution,Freese:2021zne,forte2022parton,Yu:2024ovn}, thereby providing insights into the complexities of the system. TMDs encapsulate crucial information regarding the three-dimensional internal structure of hadrons, particularly the spin-orbit correlations among quarks contained within them~\cite{Sivers:1989cc,Boer:1997nt,Barone:2001sp,Collins:2003fm}.

FFs encapsulate fundamental information regarding the extended structure of hadrons, as they represent matrix elements of conserved currents between hadronic states.
The electromagnetic interaction serves as a distinctive tool for probing the internal structure of the hadron. Measurements of electromagnetic FFs in both elastic and inelastic scattering, along with assessments of structure functions in deep inelastic scattering of electrons, have provided a wealth of information regarding the hadron's structure. A deficiency in precise information regarding the shapes of various form factors derived from the first principles of Quantum Chromodynamics (QCD) has been, and continues to be, a significant challenge in hadron physics. The electromagnetic FFs play a crucial role in elucidating nucleon structure and are essential for calculations involving the electromagnetic interactions with complex nuclei. 

The elastic electromagnetic FFs of hadrons are fundamental quantities that represent the probability of a hadron absorbing a virtual photon with four-momentum squared $Q^2$. These FFs serve as essential tools for investigating the dynamics of strong interactions across a broad spectrum of momentum transfers~\cite{Gao:2003ag,Hyde:2004gef}. Their comprehensive understanding is crucial for elucidating various aspects of both perturbative and nonperturbative hadron structure. At high momentum transfers, they convey information regarding the quark substructure of the nucleon as described by QCD. Conversely, at low momentum transfers, these quantities are influenced by fundamental properties of the nucleon, such as its charge and magnetic moment. The FFs also provide crucial insights into nucleon radii and the coupling constants of vector mesons.

The electromagnetic structure of the spin-1 $\rho$ meson as revealed in elastic electron–hadron scattering is parametrized in terms of the three form factors, the charge $G_C$, magnetic $G_M$, and quadrupole $G_Q$ form factors. The comprehension of these form factors is critically important in any theoretical framework or model pertaining to strong interactions.

The work presented here is an update and extension of our previous paper~\cite{Zhang:2022zim}. As in that paper, we study the $5$ unpolarized and $4$ polarized GPDs in the Nambu--Jona-Lasinio (NJL) model~\cite{RevModPhys.64.649,Buballa:2003qv,Zhang:2018ouu,Zhang:2016zto,Cui:2017ilj,Cui:2016zqp,Zhang:2024dhs}. Through the GPDs, we study the form factors of $\rho$ meson, which relate to the Mellin moments of GPDs. In this paper, we study the charge $G_C$, magnetic $G_M$, and quadrupole $G_Q$ form factors of $\rho$ meson and compare with the lattice data in Refs.~\cite{QCDSF:2008tjq,Shultz:2015pfa}. Through the three form factors we studied the structure functions $A(Q^2)$, $B(Q^2)$, and tensor tensor polarization function $T_{20}(Q^2,\theta)$ appeared in Rosenbluth cross section for elastic electron scattering. In addition, we study the parton distribution function of $\rho$ meson in impact parameter space. The diagrams of \( x \cdot q_C(x, \bm{b}_{\perp}^2) \), \( x \cdot q_Q(x, \bm{b}_{\perp}^2) \), and \( x \cdot q_M(x, \bm{b}_{\perp}^2) \) for various values of \( x \) and \( b_{\perp} \) are presented. The width distribution of the three distributions in the $\rho$ meson for a given momentum fraction $x$ are studied. The width distributions of \( q_C(x, \bm{b}_{\perp}^2) \), \( q_Q(x, \bm{b}_{\perp}^2) \), and \( q_M(x, \bm{b}_{\perp}^2) \) provide insights into the sizes of the charge, magnetic moment, and quadrupole moment in impact parameter space.

This article is structured as follows: In Sec.~\ref{nice}, we introduce the NJL model and subsequently present the form factors of the $\rho$ meson derived from the Mellin moments of GPDs. In Sec.~\ref{good}, we examine the properties of $\rho$ meson PDFs in impact parameter space. A concise summary and outlook are provided in Sec.~\ref{excellent}.




\section{Form factors}\label{nice}

\subsection{NJL model}

The SU(2) flavor NJL Lagrangian is presented as follows:
\begin{align}\label{1}
\mathcal{L}&=\bar{\psi }\left(i\gamma ^{\mu }\partial _{\mu }-\hat{m}\right)\psi\nonumber\\
&+G_{\pi }\left[\left(\bar{\psi }\psi\right)^2-\left( \bar{\psi }\gamma _5 \vec{\tau }\psi \right)^2\right]-G_{\omega}\left(\bar{\psi }\gamma _{\mu}\psi\right)^2\nonumber\\
&-G_{\rho}\left[\left(\bar{\psi }\gamma _{\mu} \vec{\tau } \psi\right)^2+\left( \bar{\psi }\gamma _{\mu}\gamma _5 \vec{\tau } \psi \right)^2\right],
\end{align}
the expression $\hat{m}=\text{diag}\left(m_u,m_d\right)$ denotes the current quark mass matrix, while $\vec{\tau}$ represents the Pauli matrices used to describe isospin. In the scenario of perfect isospin symmetry, we have $m_u = m_d = m$. The 4-fermion coupling constants associated with each chiral channel are designated as $G_{\pi}$, $G_{\omega}$, and $G_{\rho}$.

By solving the gap equation, we derive the dressed quark propagator within the framework of the NJL model,
\begin{align}\label{2}
S(k)=\frac{1}{{\not\!k}-M+i \varepsilon}.
\end{align}

The interaction kernel of the gap equation is local, leading us to derive a constant dressed quark mass $M$, which satisfies the following condition:
\begin{align}\label{3}
M=m+12 i G_{\pi}\int \frac{\mathrm{d}^4l}{(2 \pi )^4}\text{tr}_\text{D}[S(l)],
\end{align}
where the trace is over Dirac indices. When the coupling constant exceeds a critical threshold, specifically when $G_{\pi} > G_{critical}$, dynamical chiral symmetry breaking can occur, leading to a nontrivial solution where $M > 0$.

The NJL model is non-renormalizable, necessitating the application of a regularization method to fully define the framework. In this context, we adopt the proper time regularization (PTR) scheme~\cite{Ebert:1996vx,Hellstern:1997nv,Bentz:2001vc}.
\begin{align}\label{4}
\frac{1}{X^n}&=\frac{1}{(n-1)!}\int_0^{\infty}\mathrm{d}\tau\, \tau^{n-1}e^{-\tau X}\nonumber\\
& \rightarrow \frac{1}{(n-1)!} \int_{1/\Lambda_{\text{UV}}^2}^{1/\Lambda_{\text{IR}}^2}\mathrm{d}\tau\, \tau^{n-1}e^{-\tau X}
\end{align}
the expression $X$ represents a product of propagators that have been combined through Feynman parametrization. The infrared cutoff is expected to be on the order of $\Lambda_{\text{QCD}}$, and we select $\Lambda_{\text{IR}} = 0.240$ GeV. The parameters utilized in this study, including the coupling strength $G_{\pi}$, the momentum cutoff $\Lambda_{\text{UV}}$, and the current quark mass $m$, are determined through the Gell-Mann–Oakes–Renner (GMOR) relation given by
\begin{align}\label{4}
f_{\pi}^2m_{\pi}^2=-m\langle\bar{\psi}\psi \rangle,
\end{align}
and gap equation
\begin{align}\label{4}
M=m-2G_{\pi}\langle\bar{\psi}\psi \rangle,
\end{align}
where $m_{\pi}=0.140$ GeV represents the physical pion mass, $f_{\pi}=0.092$ GeV denotes the pion decay constant, the current quark mass is given by $m=0.016$ GeV, the ultraviolet cut off $\Lambda_{\text{UV}} = 0.645$ GeV and the constituent quark mass is specified as $M=0.4$ GeV, Additionally, $\langle\bar{\psi}\psi \rangle$ refers to the two-quark condensate derived from QCD sum rules. The pseudoscalar bubble diagram
\begin{align}\label{ab35}
\Pi_{PP}(Q^2)
=-\frac{3 }{2\pi ^2} \int_0^1\mathrm{d}x\,\mathcal{C}_0(M^2)+\frac{3 }{4\pi ^2} \int_0^1\mathrm{d}x\, Q^2 \bar{\mathcal{C}}_1(\sigma_1),
\end{align}
where $\sigma_1=M^2+x(1-x)Q^2$, $\mathcal{C}_0$ and $\bar{\mathcal{C}}_1$ are defined in Eq. (\ref{cfun}) of the appendix. The mass of the pion aligns with the value derived from the pole condition given by $1+2G_{\pi} \Pi_{PP} (-m_{\pi}^2)=0$. The coupling constants $G_{\omega}$ and $G_{\rho}$ are determined using the masses $m_{\omega}=0.782$ GeV and $m_{\rho}=0.770$ GeV through the relation \(1+2G_i \Pi_{VV} (-m_i^2)=0\), where \(i=(\omega,\rho)\). Here, $\Pi_{VV} (Q^2)$ represents the vector bubble diagram as defined in Eq. (\ref{vvbu}). The parameters utilized in this study are presented in Table \ref{tb1}.

The quark charge operator in the NJL model is
\begin{align}\label{bqp}
&\hat{Q}=\left(
\begin{array}{cc}
e_u & 0 \\
0 & e_d \\
\end{array}
\right)=(\frac{1}{6}+\frac{\tau_3}{2}),
\end{align}
where \( e_u \) and \( e_d \) represent the electric charges of the up and down quarks, respectively. This indicates that the quark-photon vertex possesses both isoscalar and isovector components. Consequently, the dressed quark-photon vertex can be articulated as follows:
\begin{align}\label{dqpver}
&\Lambda_{\gamma Q}^{\mu}(p',p)=\frac{1}{6}\Lambda_{\omega}^{\mu}(p',p)+\frac{\tau_3}{2}\Lambda_{\rho}^{\mu}(p',p),
\end{align}
the effective vertex
\begin{align}\label{dqp}
\Lambda_i^{\mu}(Q^2)&=\gamma^{\mu}P_{1i}(Q^2)+\frac{\sigma^{\mu\nu}q_{\nu}}{2M}P_{2i}(Q^2),
\end{align}
where $i=(\omega,\rho)$. For a point-like quark, we have \( P_{1i}(Q^2) = 1 \) and \( P_{2i}(Q^2) = 0 \). The inhomogeneous Bethe-Salpeter equation (BSE) governing the quark-photon vertex is expressed as follows:
\begin{align}\label{dqffff}
&\Lambda_{\gamma Q}^{\mu}(p',p)=\gamma^{\mu}(\frac{1}{6}+\frac{\tau_3}{2})\nonumber\\
&+\sum_{\Omega}K_{\Omega}\Omega\int \frac{d^4k}{(2\pi)^4}\text{tr}[\bar{\Omega} S(k+q)\Lambda_{\gamma Q}^{\mu}(k+q,k)S(k)],
\end{align}
The expression $\sum_{\Omega}K_{\Omega}\Omega_{\alpha\beta} \bar{\Omega}_{\gamma\delta}$ denotes the interaction kernels. Among these, only the isovector-vector term, $-2iG_{\rho}(\gamma_{\mu}\vec{\tau })_{\alpha\beta}(\gamma_{\mu}\vec{\tau })_{\gamma\delta}$, and the isoscalar-vector term, $-2iG_{\omega}(\gamma_{\mu})_{\alpha\beta}(\gamma_{\mu})_{\gamma\delta}$, can contribute significantly.

From the inhomogeneous BSE, the dressed quark form factors associated with the electromagnetic current described in Eq. (\ref{dqp}) are~\cite{Zhang:2021shm}
\begin{align}\label{bsam}
P_{1i}(Q^2)=\frac{1}{1+2G_i \Pi_{VV}(Q^2)},\quad \quad P_{2i}(Q^2)=0,
\end{align}
where $i=(\omega,\rho)$, $\Pi_{VV}$ is the bubble diagram
\begin{align}\label{vvbu}
\Pi_{VV}(Q^2)=\frac{3 }{\pi ^2} \int_0^1\mathrm{d}x\, x (1-x) Q^2 \bar{\mathcal{C}}_1(\sigma_1).
\end{align}
We will employ the Gamma functions and the notations presented in Eq. (\ref{cfun}) in the subsequent section.

The $\rho$ meson vertex function, in the light-cone normalization, is defined as
\begin{align}\label{6C}
\Gamma_{\rho}^{\mu}=\sqrt{Z_{\rho}}\gamma^{\mu},
\end{align}
where $Z_{\rho}$ is the square of the effective meson-quark-quark coupling constant, which is defined as follows:
\begin{align}\label{ab35}
Z_{\rho}^{-1}&=-\frac{\partial}{\partial Q                                                                                                                                                                                                                                                                                                                                                                                                                                                                                                                                                                                           ^2}\Pi_{VV}(Q^2)|_{Q^2=-m_{\rho}^2}.
\end{align}

\begin{center}
\begin{table}
\caption{Parameter set used in our work. The dressed quark mass and regularization parameters are in units of GeV, while coupling constant are in units of GeV$^{-2}$, and quark condensates are in units of GeV$^3$.}\label{tb1}
\begin{tabular}{p{0.7cm} p{0.8cm} p{0.5cm} p{0.7cm}p{0.7cm}p{0.7cm}p{0.7cm}p{0.7cm}p{0.8cm}p{1.0cm}}
\hline\hline
$\Lambda_{\text{IR}}$&$\Lambda _{\text{UV}}$&$M$&$m$&$G_{\pi}$&$Z_{\rho}$ &$m_{\rho}$&$G_{\omega}$&$G_{\rho}$&$\langle\bar{u}u \rangle^{1/3}$\\
\hline
0.24&0.645&0.4&0.016&19.0& 6.96&0.77&10.4&11.0&$-$0.173\\
\hline\hline
\end{tabular}
\end{table}
\end{center}

\subsection{FFs}
\begin{figure}
\centering
\includegraphics[width=0.47\textwidth]{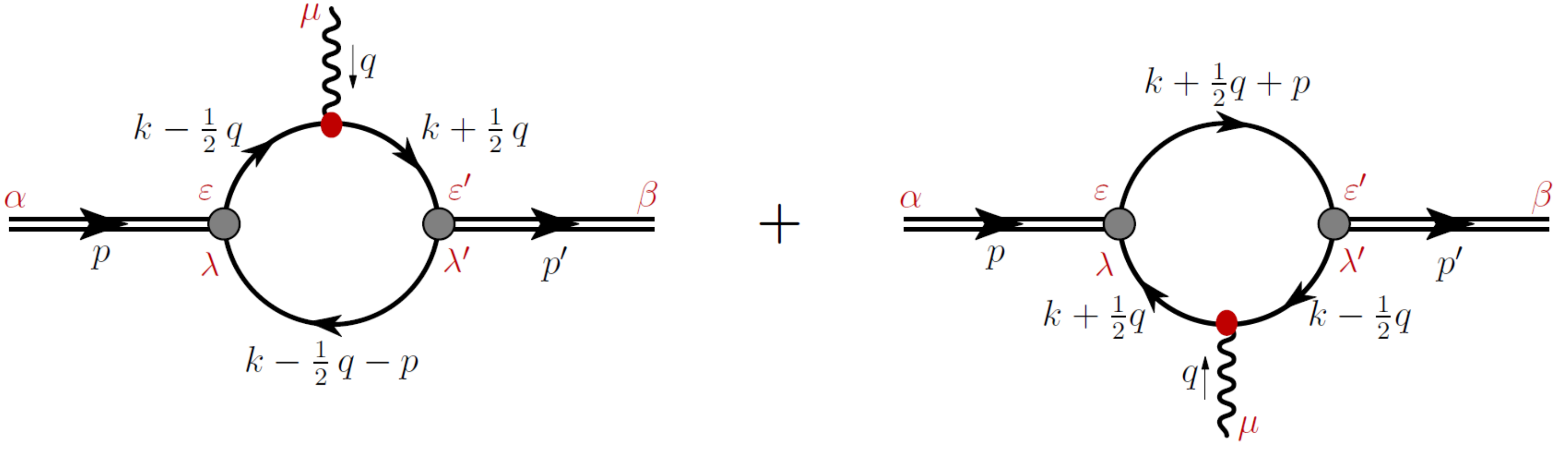}
\caption{Feynman diagrams representing the $\rho^+$ meson GPDs.}\label{GPD}
\end{figure}

The Feynman diagrams representing the GPDs of the $\rho$ meson are illustrated in Fig. \ref{GPD}. 
In the NJL model, the GPDs of the quark in the $\rho$ meson are defined as follows:
\begin{align}\label{gpddd}
V_{\mu\nu}&=2i N_c Z_{\rho}\int \frac{\mathrm{d}^4k}{(2 \pi )^4}\delta_n^x (k)\nonumber\\
&\times  \text{tr}_{\text{D}}\left[\gamma^{\mu} S \left(k_{+q}\right)\gamma^+S\left(k_{-q}\right)\gamma^{\nu} S\left(k-P\right)\right],
\end{align}
where $\text{tr}_{\text{D}}$ indicates a trace over spinor indices, $\delta_n^x (k)=\delta (xP^+-k^+)$, $k_{+q}=k+\frac{q}{2}$, $k_{-q}=k-\frac{q}{2}$. $p$ is the incoming and $p'$ the outgoing $\rho$ meson momentum, in this paper we will use the symmetry notation, the kinematics of this process and the related quantities are defined as
\begin{align}\label{4}
p^2=p^{'2}=m_{\rho}^2, \quad \quad t=q^2=(p'-p)^2=-Q^2,
\end{align}
\begin{align}\label{5}
\xi=\frac{p^+-p'^+}{p^++p'^+},\quad P=\frac{p+p'}{2},\quad n^2=0,
\end{align}
$\xi$ is the skewness parameter, in the light-cone coordinate
\begin{align}\label{4A}
v^{\pm}=(v^0\pm v^3), \quad  \mathbf{v}=(v^1,v^2),
\end{align}
for any four-vector, $n$ is the light-cone four-vector defined as $n=(1,0,0,-1)$, in the light-cone coordinate
\begin{align}\label{4B}
v^+=v\cdot n.
\end{align}
The vector quark correlator can be decomposed as follows:
\begin{align}\label{aX1}
V_{\mu\nu}&=-g^{\mu\nu}H_1+\frac{n^{\mu}P^{\nu}+P^{\mu}n^{\nu}}{n\cdot P}H_2-\frac{2P^{\mu}P^{\nu}}{m_{\rho}^2}H_3\nonumber\\
&+\frac{n^{\mu}P^{\nu}-P^{\mu}n^{\nu}}{n\cdot P}H_4+\left[\frac{m_{\rho}^2n^{\mu}n^{\nu}}{(n\cdot P)^2}+\frac{1}{3}g^{\mu\nu}\right]H_5,
\end{align}
the expressions for \( H_i\) are derived in Ref.~\cite{Zhang:2022zim}. The general form of the vector current associated with the \(\rho\) meson is presented as follows:
\begin{align}\label{bF3}
j_{\rho}^{\alpha,\mu\nu }&=[-g^{\mu\nu }F_1(t)-\frac{2P^{\mu}P^{\nu }}{m_{\rho}^2}F_3(t)](p^{\alpha}+p^{'\alpha})\nonumber\\
&+2(P^{\nu}g^{\alpha \mu}+P^{\mu}g^{\alpha\nu})F_2(t),
\end{align}
integrating over $x$ allows one to derive
\begin{subequations}\label{bF31}
\begin{align}
\int_{-1}^1\mathrm{d}x H_i(x,\xi,t)&=F_i(t), \quad \quad (i=1,2,3) \,, \\
\int_{-1}^1\mathrm{d}x H_i(x,\xi,t)&=0, \quad \quad (i=4,5),
\end{align}
\end{subequations}
the expressions for \( F_1(Q^2) \), \( F_2(Q^2) \), and \( F_3(Q^2) \) are presented in Ref.~\cite{Zhang:2022zim}.

The Sachs-like charge, magnetic, and quadrupole form factors for the $\rho$ meson are presented as follows:
\begin{subequations}\label{sff}
\begin{align}
G_C(Q^2)&=F_1(Q^2)+\frac{2}{3}\eta \,G_Q(Q^2)\,, \\
G_M(Q^2)&=F_2(Q^2)\,, \\
G_Q(Q^2)&=F_1(Q^2)+(1+\eta)F_3(Q^2)-F_2(Q^2),
\end{align}
\end{subequations}
where $\eta=Q^2/(4m_{\rho}^2)$. The dressed Sachs-like charge, magnetic, and quadrupole form factors are defined as $G_C^D(Q^2)=G_C(Q^2)P_{1\rho}(Q^2)$, $G_M^D(Q^2)=G_M(Q^2)P_{1\rho}(Q^2)$, and $G_Q^D(Q^2)=G_Q(Q^2)P_{1\rho}(Q^2)$, respectively. Here, \( P_{1\rho}(Q^2) \) is specified in Eq. (\ref{bsam}).
\begin{figure}
\centering
\includegraphics[width=0.47\textwidth]{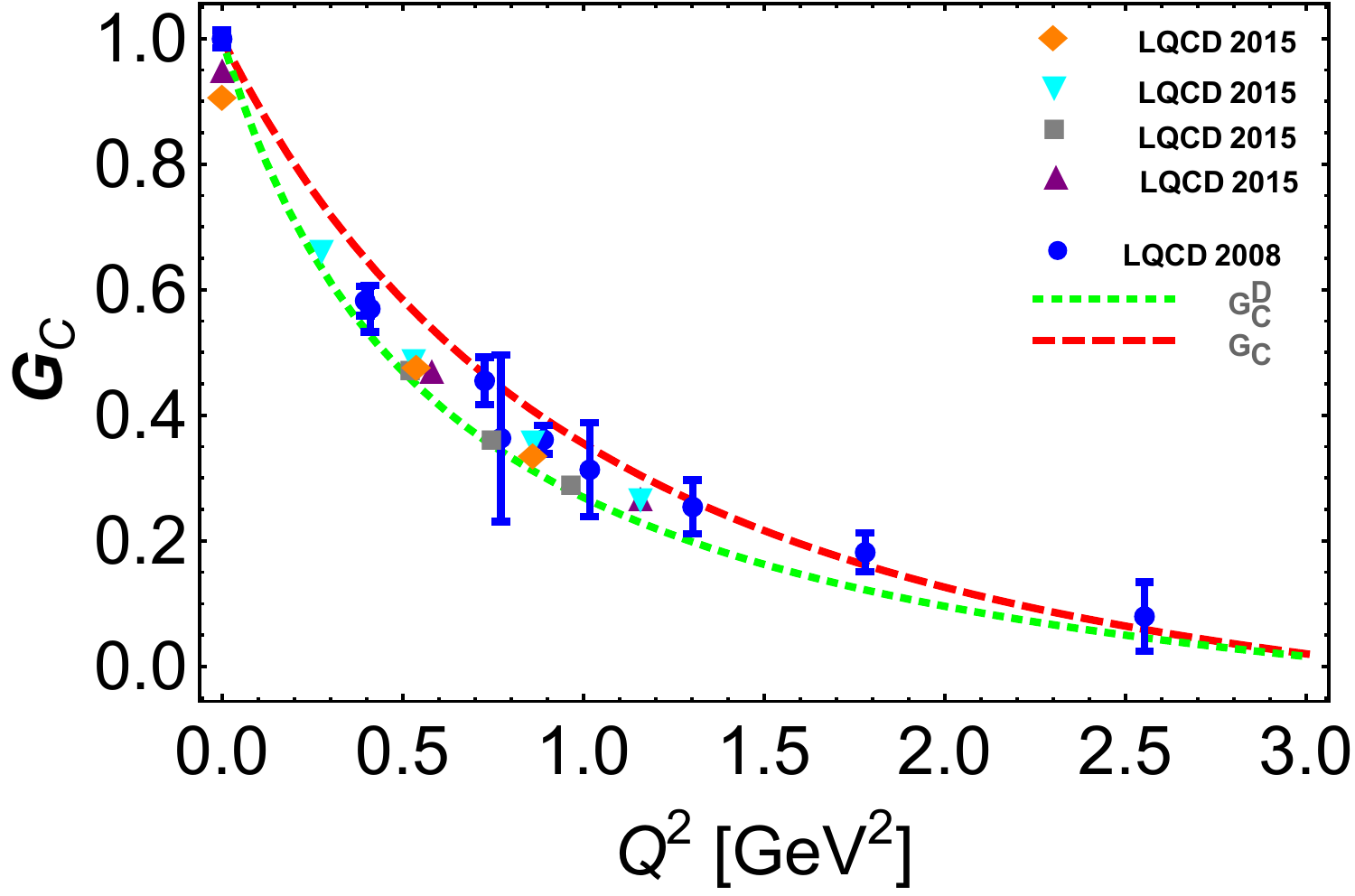}
\qquad
\includegraphics[width=0.47\textwidth]{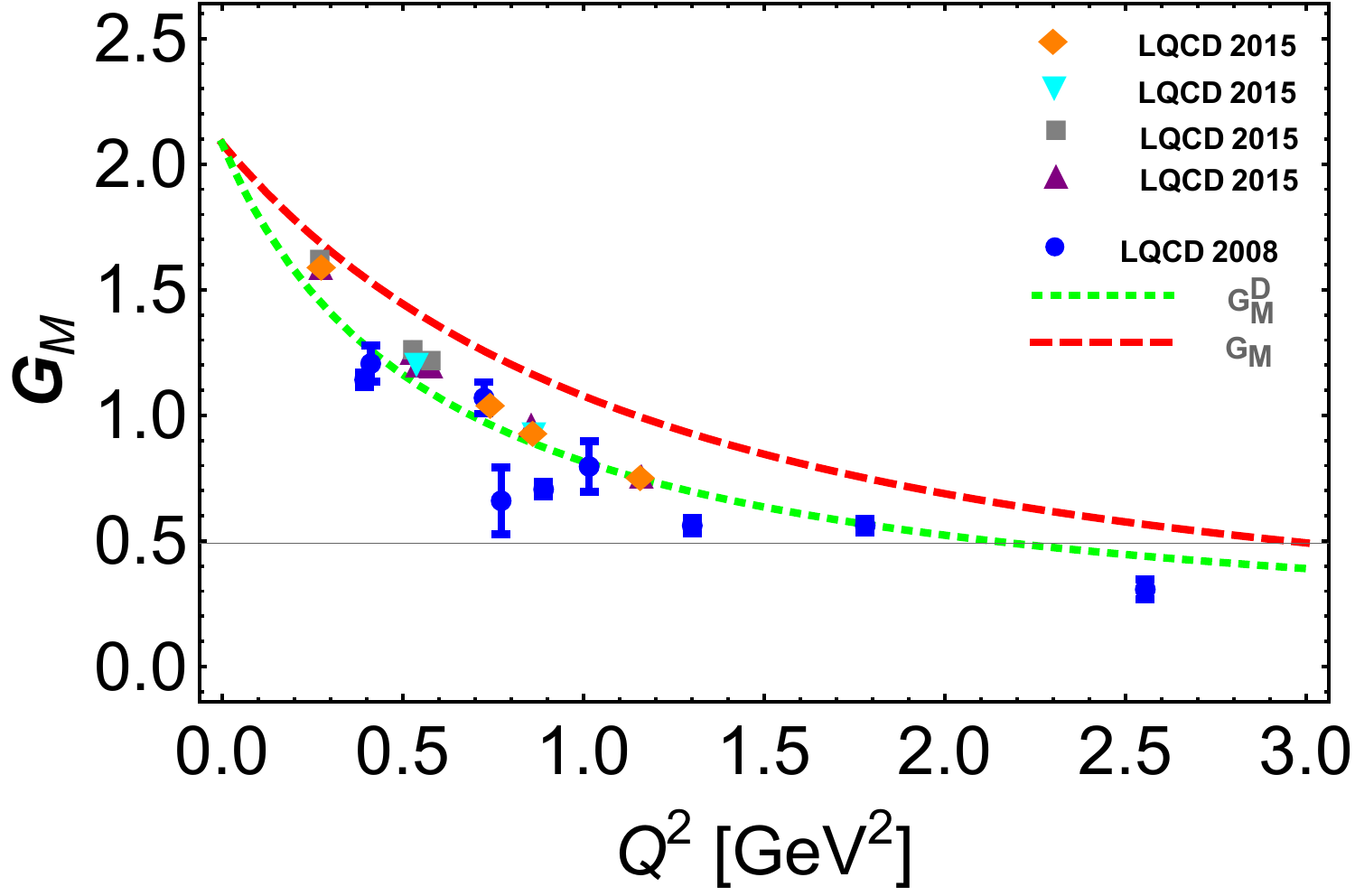}
\qquad
\includegraphics[width=0.47\textwidth]{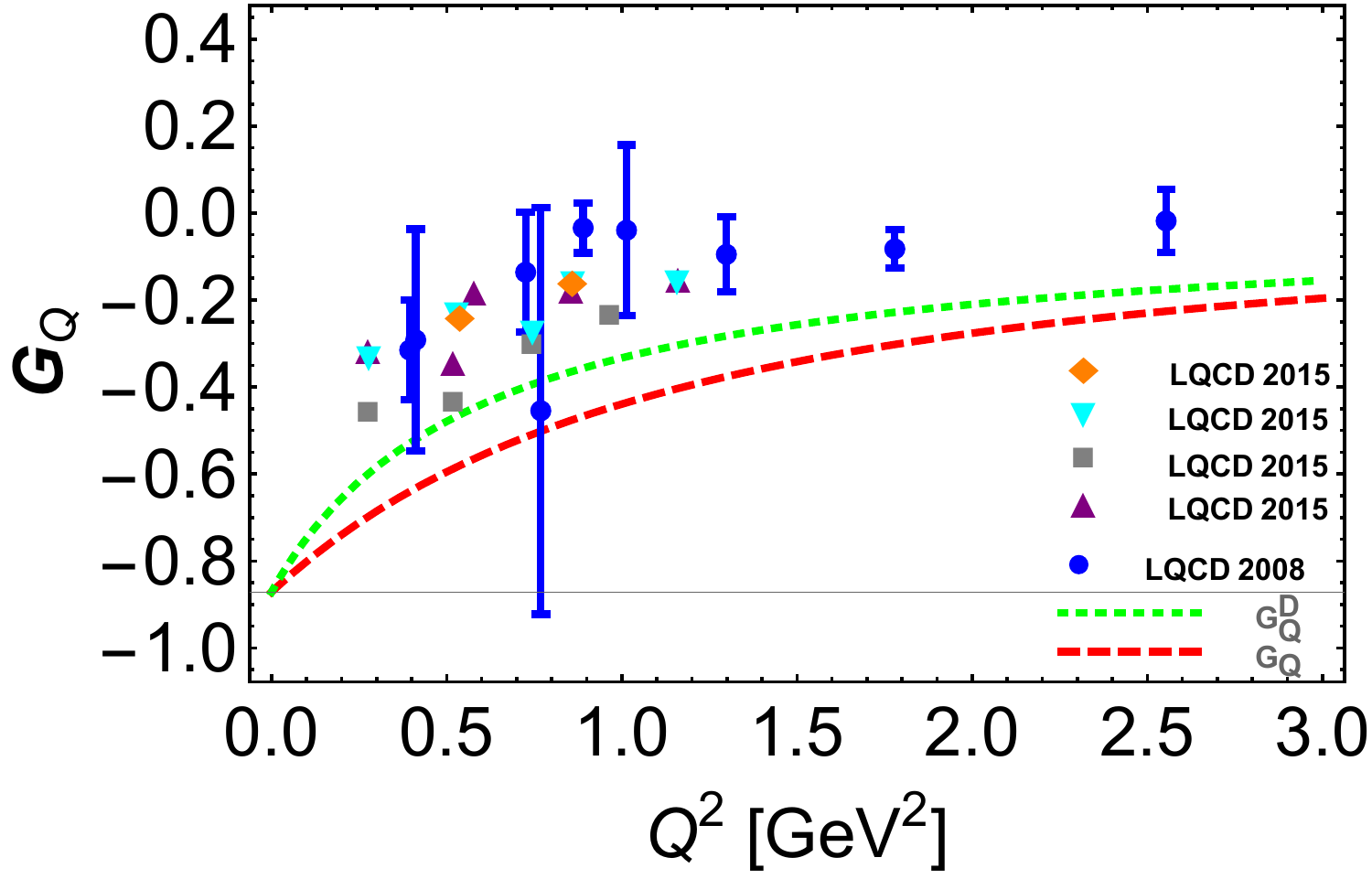}
\caption{The bare and dressed charge,  magnetic, and quadrupole from factors in the upper, middle, and lower panel, separately. The bare and the dressed form factors are denoted by the red dashed line and the green dotted line, respectively. The LQCD predictions from 2008~\cite{QCDSF:2008tjq} are denoted by blue solid lines accompanied by error bars. Additionally, LQCD predictions from 2015~\cite{Shultz:2015pfa} are represented by yellow diamonds, cyan inverted triangle, gray squares, and purple triangles. }\label{gcmq}
\end{figure}

We compare our results with the lattice QCD (LQCD) findings presented in Refs.~\cite{QCDSF:2008tjq,Shultz:2015pfa} in Fig. \ref{gcmq}. The figure illustrates that the general trends of both the bare and dressed charge factor are consistent with the lattice results. The error bars of the 2015 lattice data in Ref.~\cite{Shultz:2015pfa} are quite short, particularly at large values of \(Q^2\), which results in them appearing as dots when plotted. The various points without uncertainties for the same kinematic conditions represent different initial and final $\rho$ meson momenta. It is important to note that form factors corresponding to different initial and final $\rho$ meson momenta at the same \(Q^2\) may overlap.

Regarding the magnetic form factor, as shown in the middle diagram, we observe that the bare $G_M$ is harder than what is reported by lattice studies; conversely, the dressed $G_M^D$ aligns more closely with the lattice results from~\cite{Shultz:2015pfa}, although it remains harder than those from~\cite{QCDSF:2008tjq}.

For the quadrupole form factor $G_Q$, both bare and dressed form factors exhibit a hardness greater than that observed in Refs.~\cite{QCDSF:2008tjq,Shultz:2015pfa}.

\begin{figure}
\centering
\includegraphics[width=0.47\textwidth]{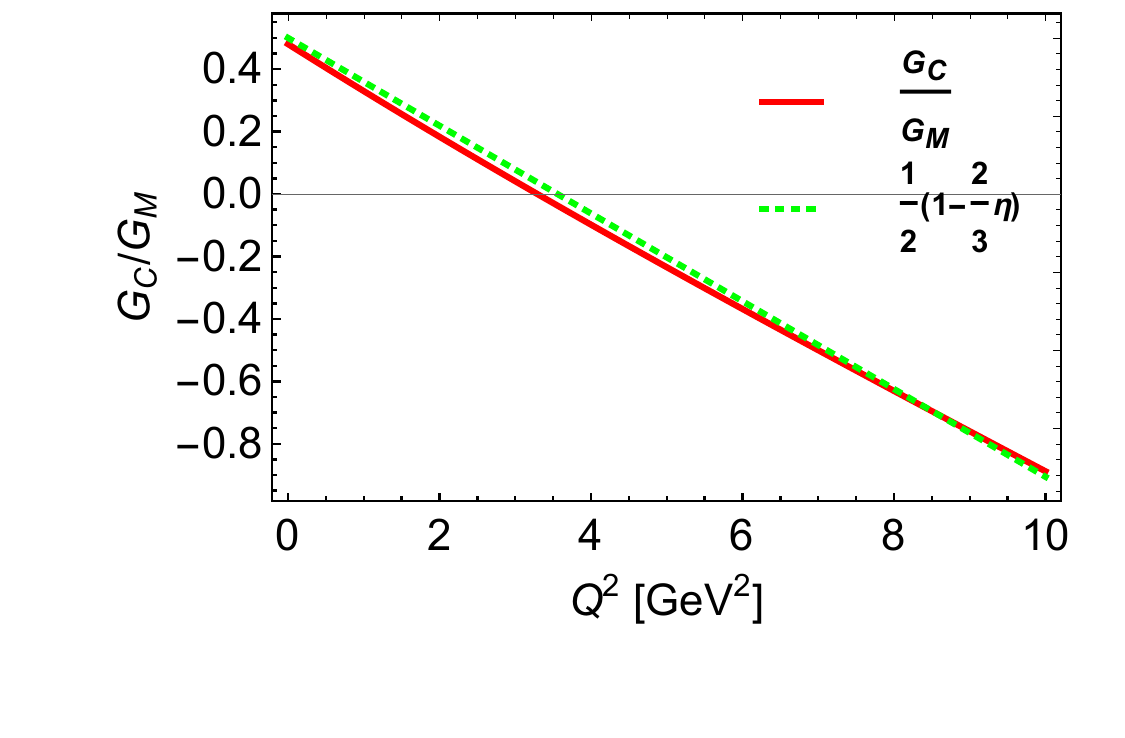}
\qquad
\includegraphics[width=0.47\textwidth]{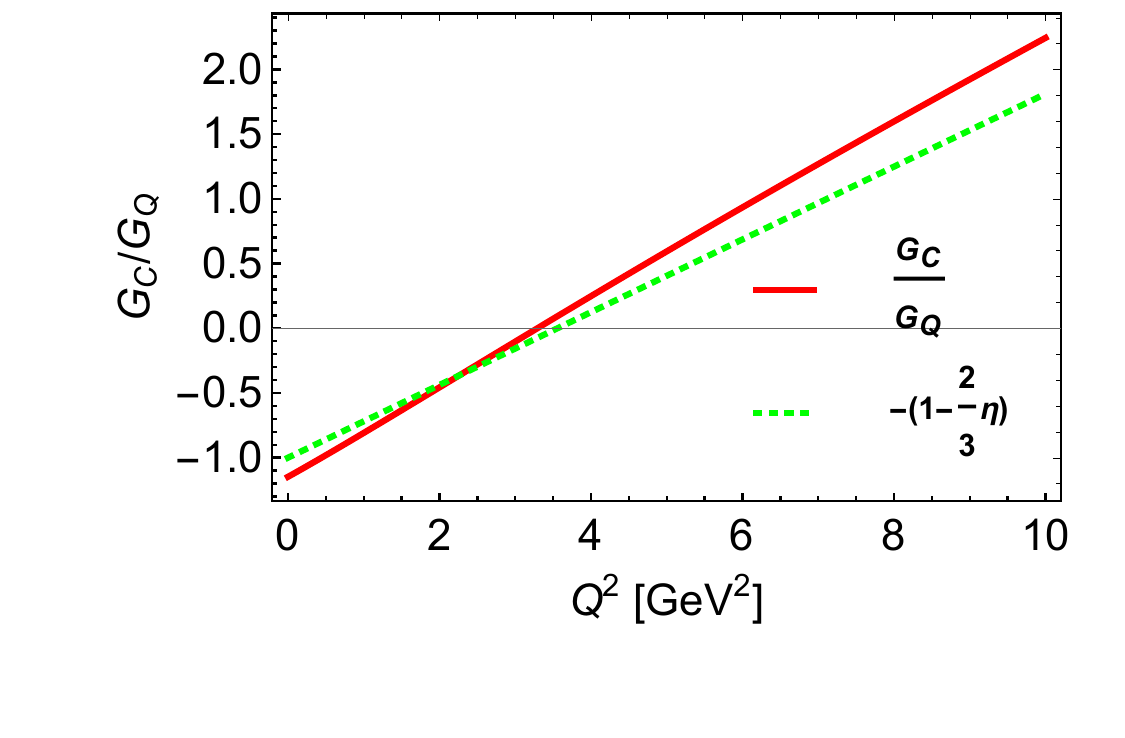}
\qquad
\includegraphics[width=0.47\textwidth]{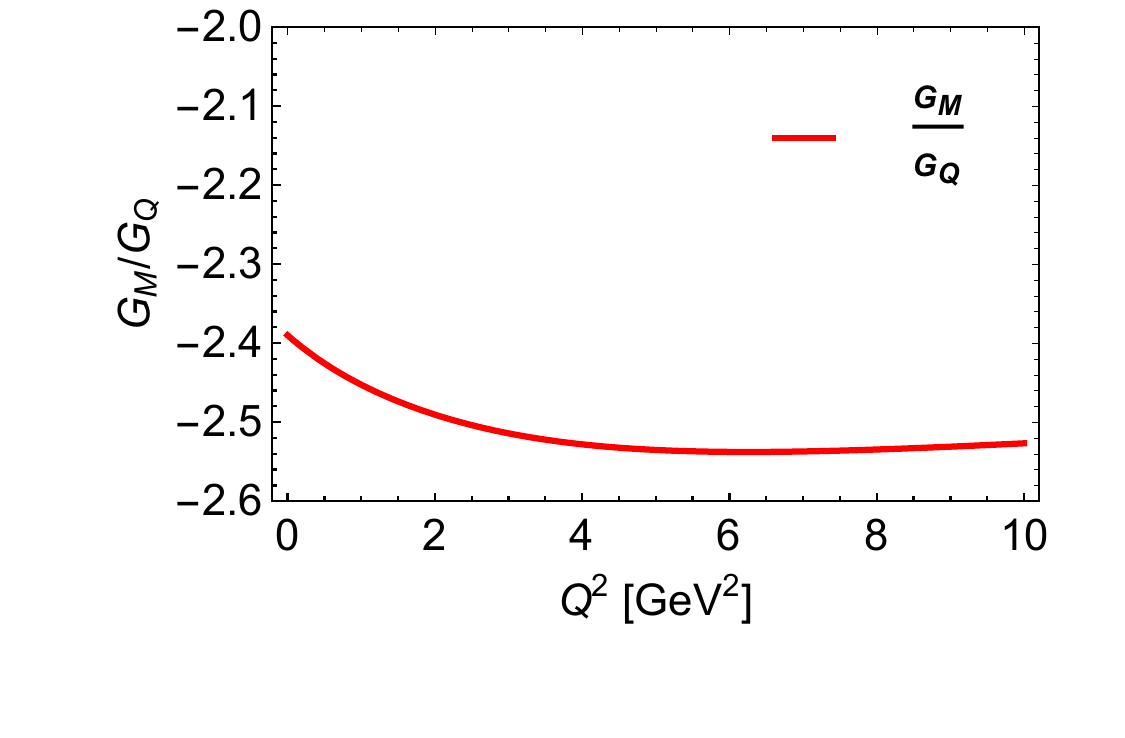}
	\caption{The upper panel: The red solid line represents the ratio of \( G_C/G_M \), while the green dotted line denotes \(\frac{1}{2}(1-\frac{2}{3}\eta) \). The middle panel: The red solid line illustrates the ratio of \( G_C/G_Q \), and the green dotted line indicates \(-(1-\frac{2}{3}\eta) \). The lower panel: The red solid line depicts the ratio of \( G_M/G_Q \).}\label{ratio}
\end{figure}

In Ref.~\cite{Brodsky:1992px}, a relationship was derived for the form factors of spin-$1$ particles at large $Q^2$. Specifically, in the regime of large timelike or spacelike momenta, the ratio of form factors for the $\rho$ meson is expected to exhibit behavior as follows:
\begin{align}\label{lsff}
G_C(Q^2):G_M(Q^2):G_Q(Q^2)=(1-\frac{2}{3}\eta):2:-1
\end{align}
where the corrections are of the orders $\Lambda_{\text{QCD}}/Q$ and $\Lambda_{\text{QCD}}/m_{\rho}$. In Fig. \ref{ratio}, we present a diagram illustrating the ratios of the three form factors. From these figures, it is evident that the ratios $G_C/G_M$ and $G_C/G_Q$ exhibit a good fit, while the ratio $G_M/G_Q$ remains below $-2$ and approaches a finite value of $-2.52$ as $Q^2$ increases. 
%

\begin{figure*}
\centering
\includegraphics[width=0.47\textwidth]{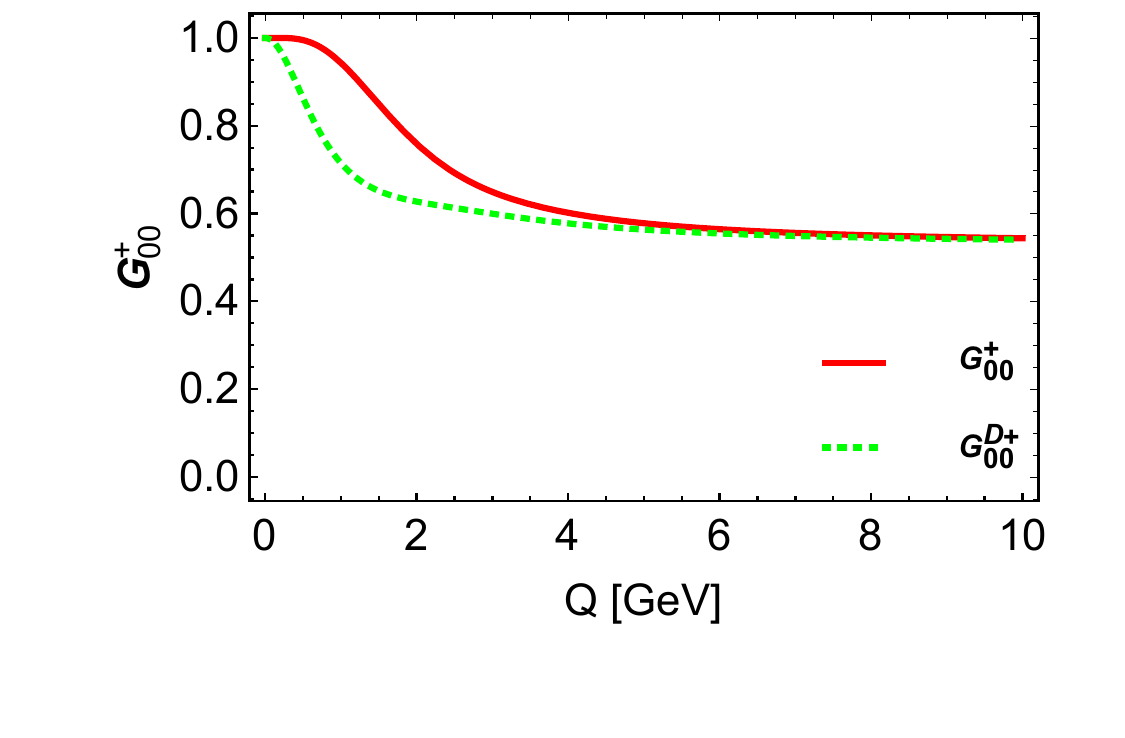}
\qquad
\includegraphics[width=0.47\textwidth]{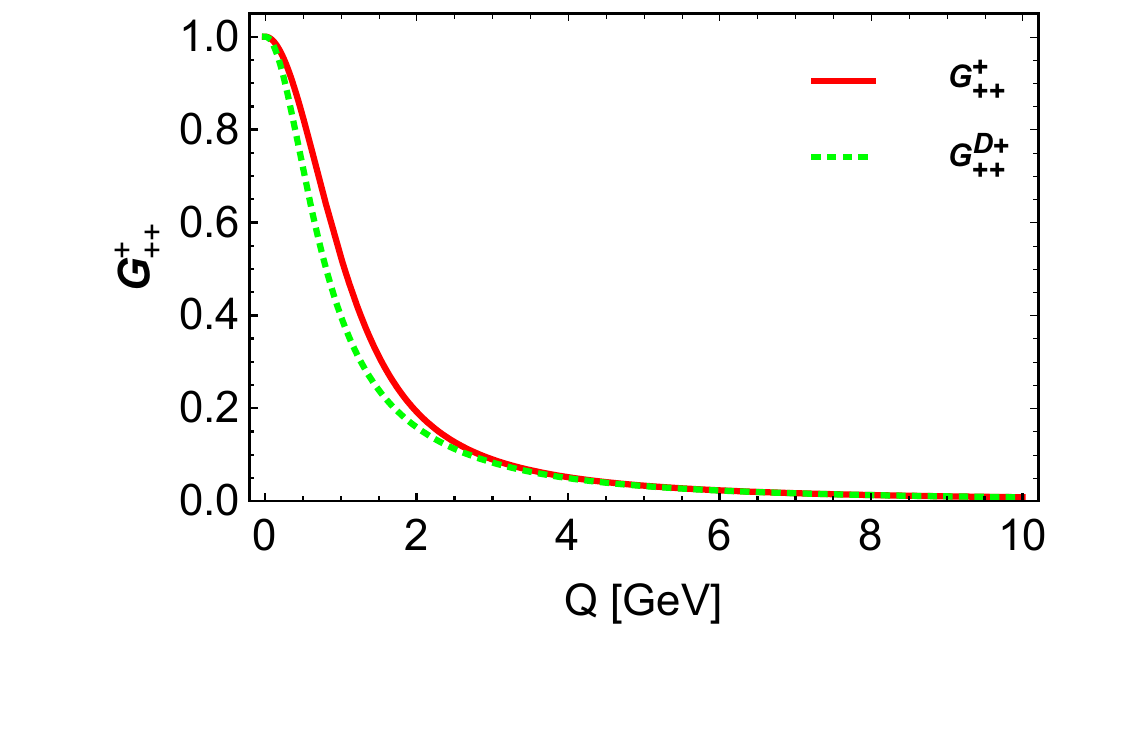}
\qquad
\includegraphics[width=0.47\textwidth]{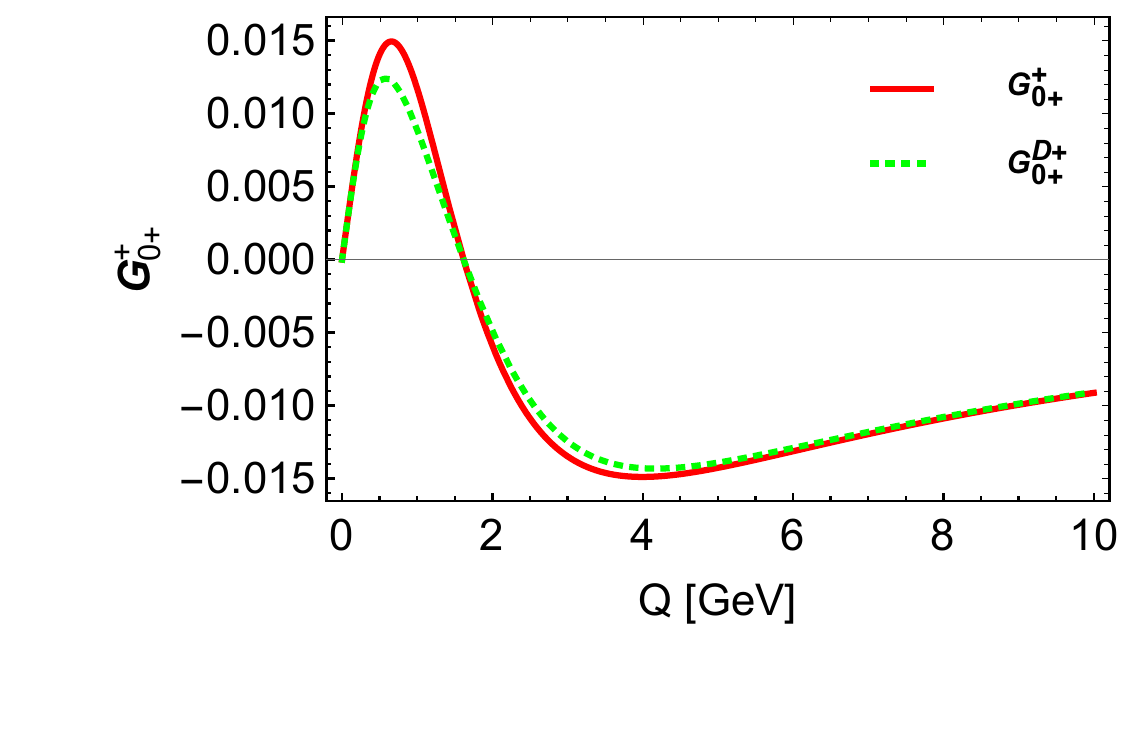}
\qquad
\includegraphics[width=0.47\textwidth]{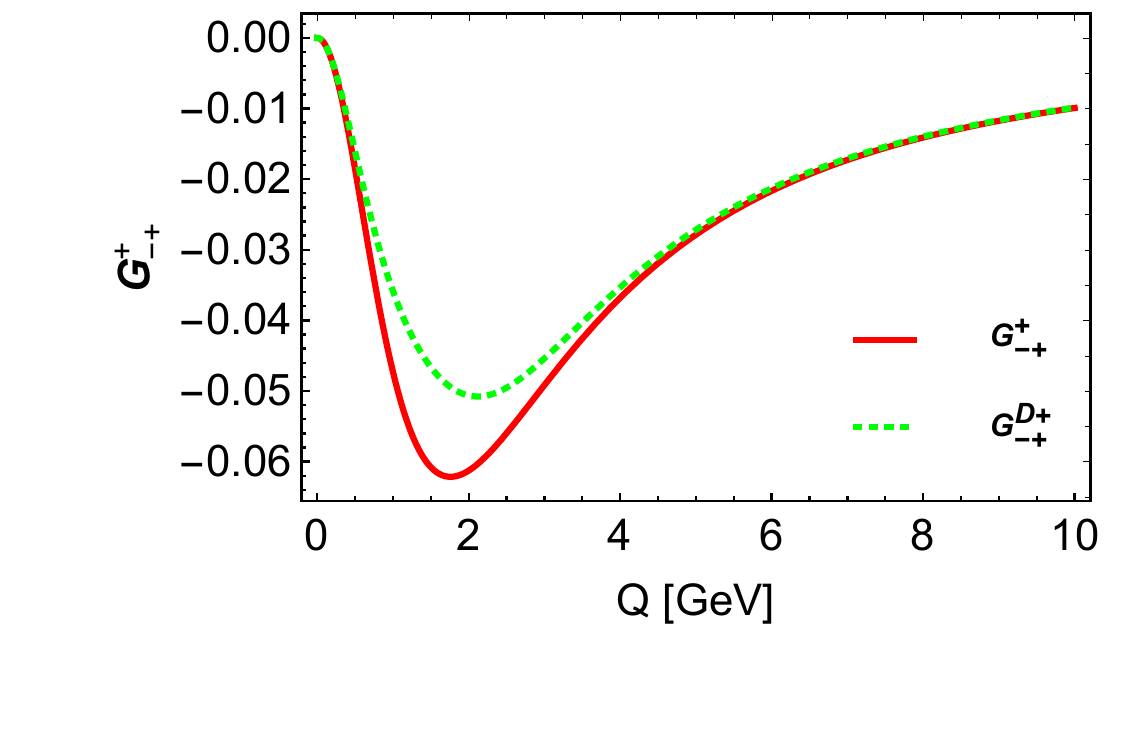}
\caption{The red solid line represents the helicity-conserving and helicity non-conserving matrix elements, which are denoted as \( G_{00}^+ \), \( G_{11}^+ \), \( G_{0+}^+ \), and \( G_{-+}^+ \). The green dotted line illustrates the dressed helicity-conserving and helicity non-conserving matrix elements, indicated as \( G_{00}^{D+} \), \( G_{11}^{D+} \), \( G_{0+}^{D+} \), and \( G_{-+}^{D+} \).}\label{gcnc}
\end{figure*}

Further, one can define the helicity-conserving matrix elements $(G_{11}^+,G_{00}^+)$ and helicity nonconserving matrix elements $(G_{0+}^+,G_{-+}^+)$, respectively, in terms of $G_C$, $G_M$, $G_C$ as Refs.~\cite{Ramalho:2023hqd,Kumar:2019eck,Mondal:2017lph}
\begin{subequations}\label{g00}
\begin{align}
G_{11}^+&=\frac{1}{1+\eta}\left(G_C+\eta G_M+\frac{\eta}{3}G_Q\right)  \,, \\
G_{00}^+&=\frac{1}{1+\eta}\left((1-\eta)G_C+2\eta G_M-\frac{2\eta}{3}(1+2\eta)G_Q\right)\,, \\
G_{0+}^+&=-\frac{\sqrt{2\eta}}{1+\eta}\left(G_C-\frac{1}{2}(1-\eta)G_M+\frac{\eta}{3}G_Q\right)\,, \\
G_{-+}^+&=\frac{\eta}{1+\eta}\left(G_C-G_M-(1+\frac{2\eta}{3})G_Q\right).
\end{align}
\end{subequations}
In Fig. \ref{gcnc}, we present the diagrams for both the bare and dressed helicity-conserving matrix elements $(G_{11}^+, G_{00}^+)$, as well as the helicity non-conserving matrix elements $(G_{0+}^+, G_{-+}^+)$. For $G_{00}^+$, the contact interaction renders the form factor hard; as $Q$ approaches infinity, $G_{00}^+$ approximates a non-zero constant value. The dressed version, $G_{00}^{D+}$, is softer than its bare counterpart $G_{00}^+$; when $Q > 4$ GeV, it remains nearly constant at approximately $G_{00}^{D+}\simeq 0.54$. In comparison with Ref.~\cite{Kumar:2019eck,Mondal:2017lph}, which indicates that their result for $G_{00}^+$ approaches zero as $Q$ increases, our findings suggest a harder behavior.

For the helicity-conserving element $G_{11}^+$, we observe that it also tends toward zero with increasing values of $Q$, consistent with Refs.~\cite{Kumar:2019eck,Mondal:2017lph}. 

The matrix element $G^{+}_{0+}$ displays a peak around $ Q\simeq 0.7$ GeV, aligning with observations from Ref.~\cite{Kumar:2019eck}; The concavity observed in our model occurs at approximately $ Q \simeq 3.8$ GeV. Furthermore, the peaks of the dressed $ G^{D+}_{0+}$ are smaller compared to those of $ G^{+}_{0+}$.

Regarding $G^{+}_{-+}$ , we identify a minimum value occurring at $ Q\simeq 1.8 $ GeV, which is larger than what was reported in Ref.~\cite{Kumar:2019eck}, the minimum value of $G^{+}_{-+}$ is $-0.06$. For the dressed $G_{- +}^{D+}$, a minimum value occurring at $ Q\simeq 2 $ GeV, the minimum value is $-0.05$; 

The standard Rosenbluth cross section~\cite{Hofstadter:1957wk} for elastic electron scattering on a target of arbitrary spin in the laboratory frame,
\begin{align}\label{crosss}
\frac{\mathrm{d}\sigma}{\mathrm{d}\Omega}=\frac{\alpha^2 \cos^2(\theta/2)}{4E^2\sin^2(\theta/2)}\frac{E^{'}}{E}\left[A(Q^2)+B(Q^2) \tan^2(\theta/2)\right].
\end{align}
In the context of invariants, it is:
\begin{align}\label{crosss1}
\frac{\mathrm{d}\sigma}{\mathrm{d}t}=\frac{4\pi \alpha^2}{t^2}\left[\left(1+\frac{ts}{(s-m_{\rho}^2)^2}\right)A(-t)-\frac{m_{\rho}^2t }{(s-m_{\rho}^2)^2}B(-t) \right].
\end{align}
%

\begin{figure}
\centering
\includegraphics[width=0.47\textwidth]{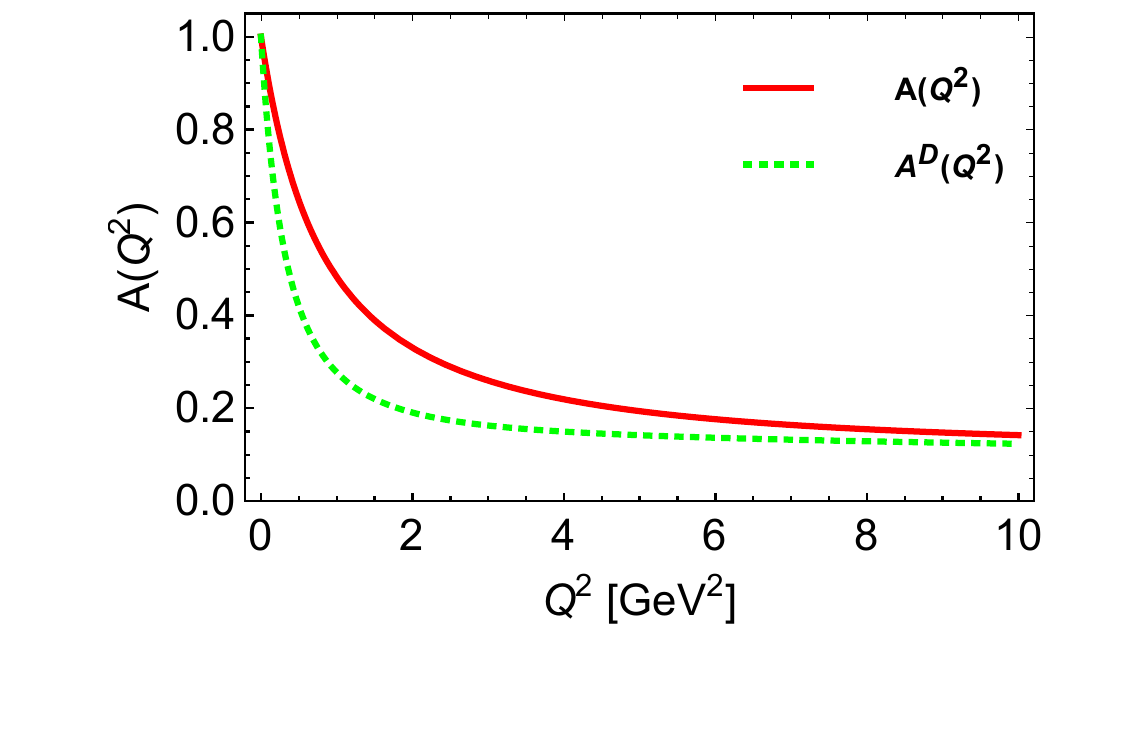}
\qquad
\includegraphics[width=0.47\textwidth]{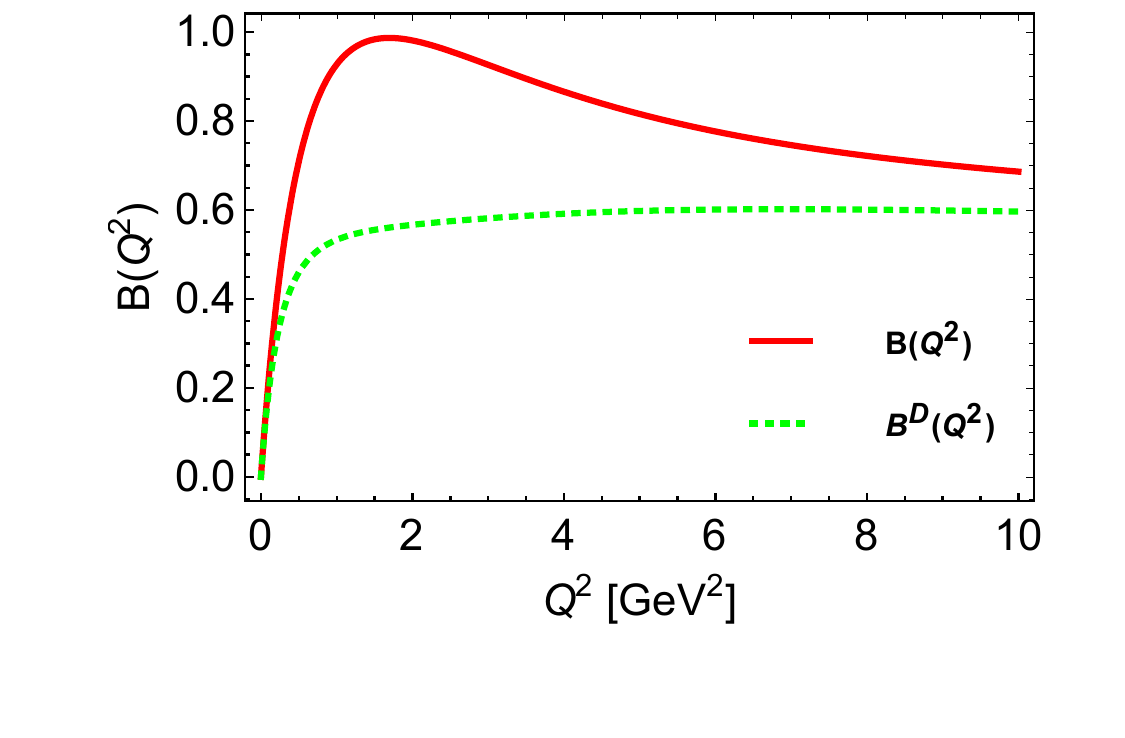}
\caption{The bare structure functions $A(Q^2)$ and $B(Q^2)$, represented by the solid red line, and the dressed structure functions $A^D(Q^2)$ and $B^D(Q^2)$, depicted by the green dotted line, of the $\rho$ meson.}\label{AB}
\end{figure}

The structure functions $A(Q^2)$, $B(Q^2)$ and the tensor polarization~\cite{Haftel:1980zz} $T_{20}(Q^2,\theta)$ are defined as,
\begin{align}\label{ab}
&A(Q^2)=G_C^2+\frac{2}{3}\eta G_M^2+\frac{8}{9}\eta^2G_Q^2\,,\\
&B(Q^2)=\frac{4}{3}\eta(1+\eta) G_M^2\,,\\
&T_{20}(Q^2,\theta)=-\eta\frac{\sqrt{2}}{3}\nonumber\\
&\times \frac{\frac{4}{3}\eta G_Q^2+4G_QG_C+(1/2+(1+\eta\tan^2\frac{\theta}{2}))G_M^2 }{A+B\tan^2\frac{\theta}{2}}.
\end{align}
In Fig \ref{AB}, we present diagrams for both the bare and dressed forms of $A(Q^2)$ and $B(Q^2)$. In our analysis, as $Q^2$ increases, $A(Q^2)$ exhibits a decreasing trend, ultimately stabilizing at a finite value — approximately $0.14$ for the bare $A(Q^2)$ and around $0.12$ for the dressed $A^D(Q^2)$.

For small values of $Q^2$, $B(Q^2)$ initially increases with rising $Q^2$, reaching a peak at approximately $Q^2\simeq 1.5$ GeV$^2$, after which it declines as $Q^2$ continues to grow. At sufficiently large values of $Q^2$, it approaches a finite limit of about $0.7$. In contrast, the dressed form, denoted as $B^D(Q^2)$, does not exhibit a maximum; instead, it remains nearly constant when $ Q^{2} > 1$ GeV$^2$, stabilizing at approximately $0.6$.

\begin{figure}
\centering
\includegraphics[width=0.47\textwidth]{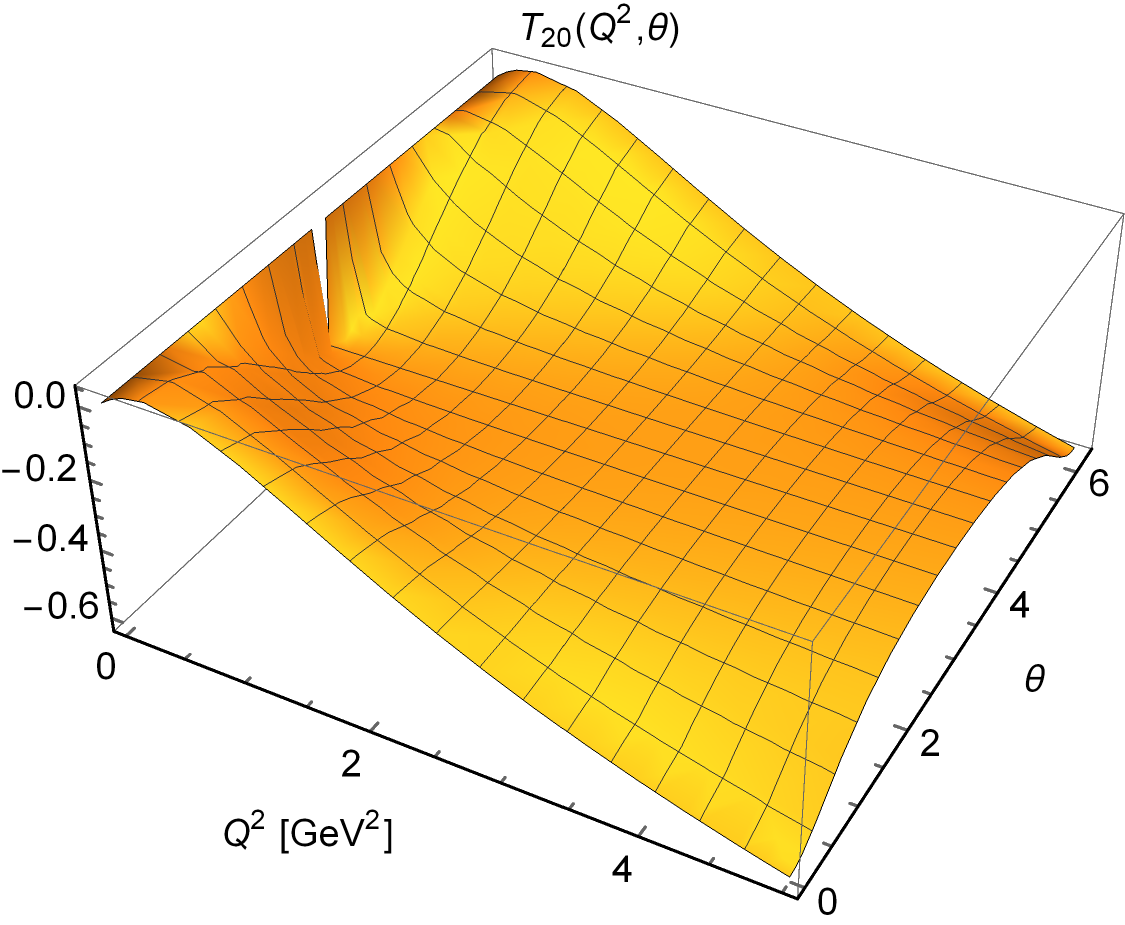}
\caption{The $\rho$ meson tensor polarization $T_{20}(Q^2,\theta)$. }\label{TQ}
\end{figure}

In Fig. \ref{TQ}, we present the three-dimensional diagram of  $T_{20}(Q^2,\theta)$. When $Q^2$ is small, $T_{20}(Q^2,\theta)$ exhibits larger values in the regions where $\theta\rightarrow 0$ and $\theta\rightarrow 2\pi$, while it is comparatively smaller near $\theta\simeq \pi$. At $\theta\simeq \pi$, as $Q^2$ increases, there is minimal change in $T_{20}(Q^2,\theta)$. Conversely, at both limits of $\theta\rightarrow 0$ and $\theta\rightarrow 2\pi$, an increase in $Q^2$, results in a gradual decrease of $T_{20}(Q^2,\theta)$. The dressed $T_{20}^D(Q^2,\theta)$ remains consistent with $T_{20}(Q^2,\theta)$.

Thus, the domain for leading-power perturbative QCD predictions regarding the $\rho$ meson form factors is characterized by $Q^2 \gg 2m_{\rho}\Lambda_{\text{QCD}} \sim 0.35$ GeV$^2$. Within this domain, one obtains
\begin{align}\label{ab1}
\frac{B}{A}&\simeq \frac{4\eta(1+\eta)}{\eta^2+\eta+3/4}\,,\\
T_{20}(\theta)&\simeq -\sqrt{2}\frac{\eta(\eta-\frac{1}{2}+(\eta+1)\tan^2\frac{\theta}{2})}{\eta^2+\eta+\frac{3}{4}+4\eta (\eta+1)\tan^2 \frac{\theta}{2}},\label{ab11}
\end{align}
In the extreme limit, $\eta\gg1$, namely, $Q^2 \gg 4m_{\rho}^2=2.37$ GeV$^2$, these reduce to
\begin{align}\label{ab2}
\frac{B}{A}&\simeq 4\,,\\
T_{20}(\theta)&\simeq -\sqrt{2}\frac{1+\tan^2\frac{\theta}{2}}{1+4\tan^2 \frac{\theta}{2}},
\end{align}
For $\eta \ll 1$, namely, $Q^2 \ll 2.37$ GeV$^2$, one obtains
\begin{align}\label{ab3}
\frac{B}{A}&\simeq \frac{16}{3}\eta \,,\\
T_{20}(\theta)&\simeq -\frac{2\sqrt{2}}{3}\eta (1-2\tan^2\frac{\theta}{2}),
\end{align}
the fundamental assumption underlying all of these findings is that the $G_{00}^+$ amplitude defined in Eq. (\ref{g00}) is predominant. 

\begin{figure}
\centering
\includegraphics[width=0.47\textwidth]{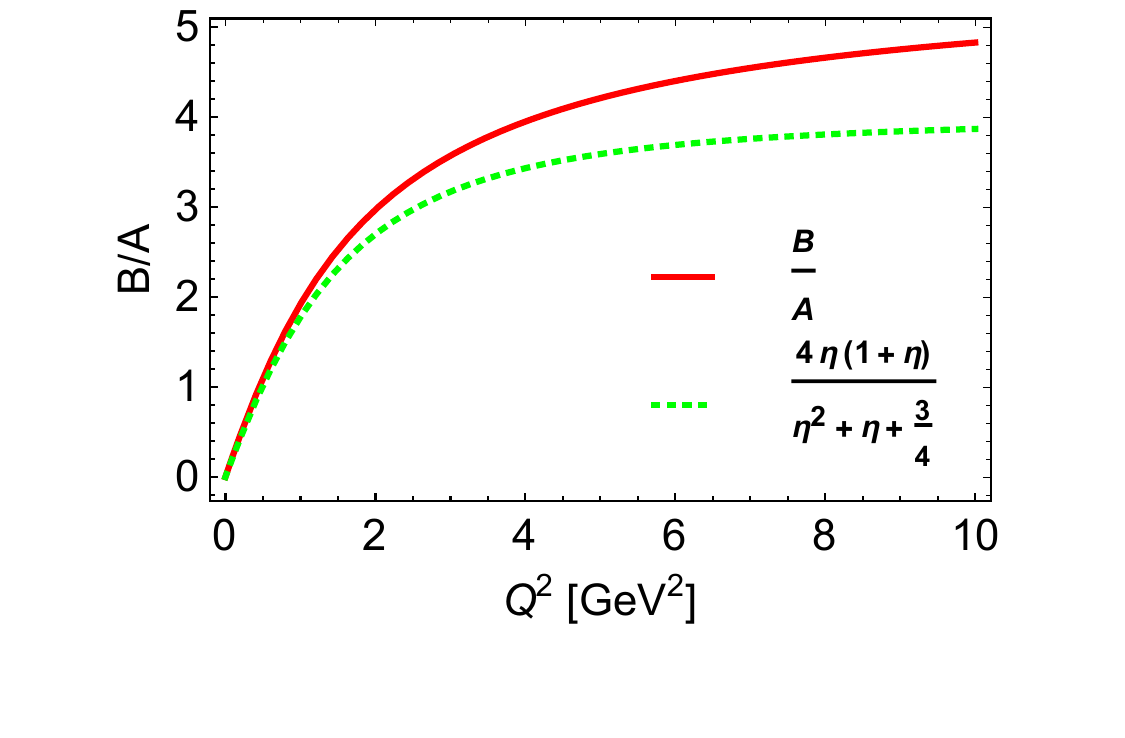}
\caption{The red solid line represents the ratio \( B/A \), while the green dotted line illustrates the expression \( \frac{4\eta(1+\eta)}{\eta^2+\eta+3/4} \) for the \(\rho\) meson.}\label{rab}
\end{figure}

In Fig. \ref{rab}, we present the ratio of $ A(Q^2)$  to $B(Q^2)$. From the diagram, it is evident that when $Q^2$ is small, the results align with theoretical predictions. As $Q^2$ increases, the value of $B/A$ exceeds $4$, approaching approximately $B/A \simeq 5$.

\begin{figure}
\centering
\includegraphics[width=0.47\textwidth]{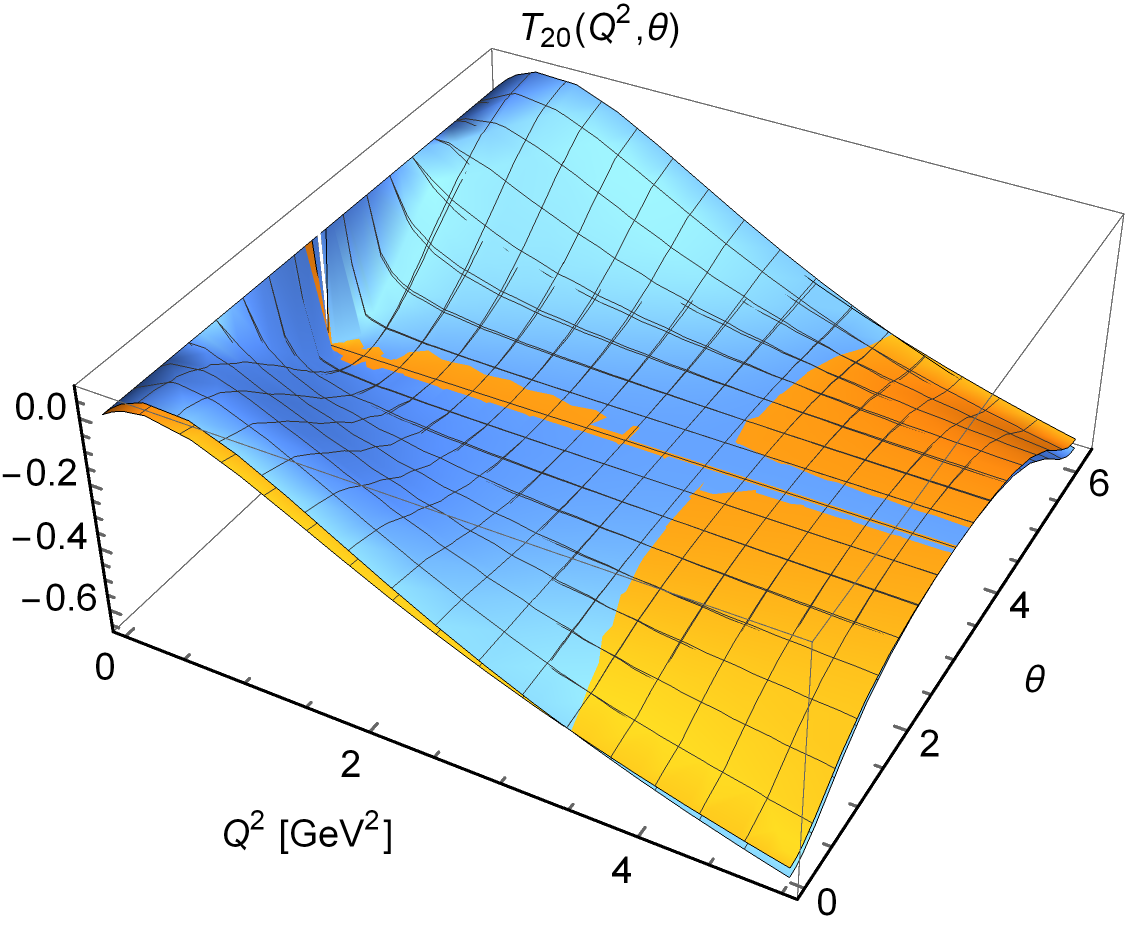}
\caption{The tensor polarization \( T_{20}(Q^2, \theta) \) is depicted in yellow, accompanied by the approximate value of the \( \rho \) meson as presented in Eq. (\ref{ab11}), which is illustrated in blue. }\label{rt}
\end{figure}

In Fig. \ref{rt}, we illustrate the graphs of $T_{20}(Q^2,\theta)$ and $T_{20}(\theta)$, as described in Eq. (\ref{ab11}). The diagrams indicate that our calculated values for $T_{20}(Q^2,\theta)$ closely coincide with those predicted by theory.
%

%
%

%
%

%
%

\section{Impact parameter dependent PDFs}\label{good}
The impact parameter dependent PDFs are defined as
\begin{align}\label{aG9}
q\left(x,\bm{b}_{\perp}^2\right)&=\int \frac{\mathrm{d}^2\bm{\Delta}_{\perp}}{(2 \pi )^2}e^{-i\bm{b}_{\perp}\cdot \bm{\Delta}_{\perp}}H\left(x,0,-\bm{\Delta}_{\perp}^2\right),
\end{align}
which means the impact parameter dependent PDFs are the Fourier transform of GPDs at $\xi=0$. We examine the quantities $q_C$, $q_M$, and $q_Q$ as discussed in Ref.~\cite{Zhang:2022zim}. In this section, we further investigate these parameters within the context of impact parameter space.

\begin{figure}
\centering
\includegraphics[width=0.47\textwidth]{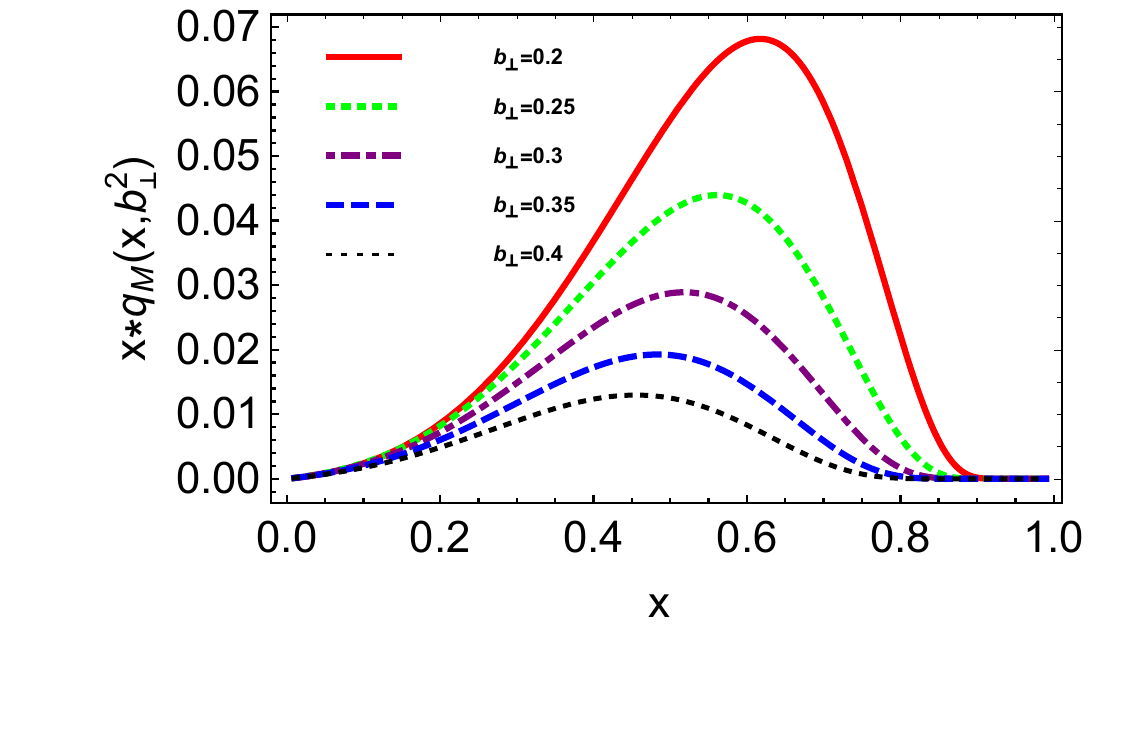}
\qquad
\includegraphics[width=0.47\textwidth]{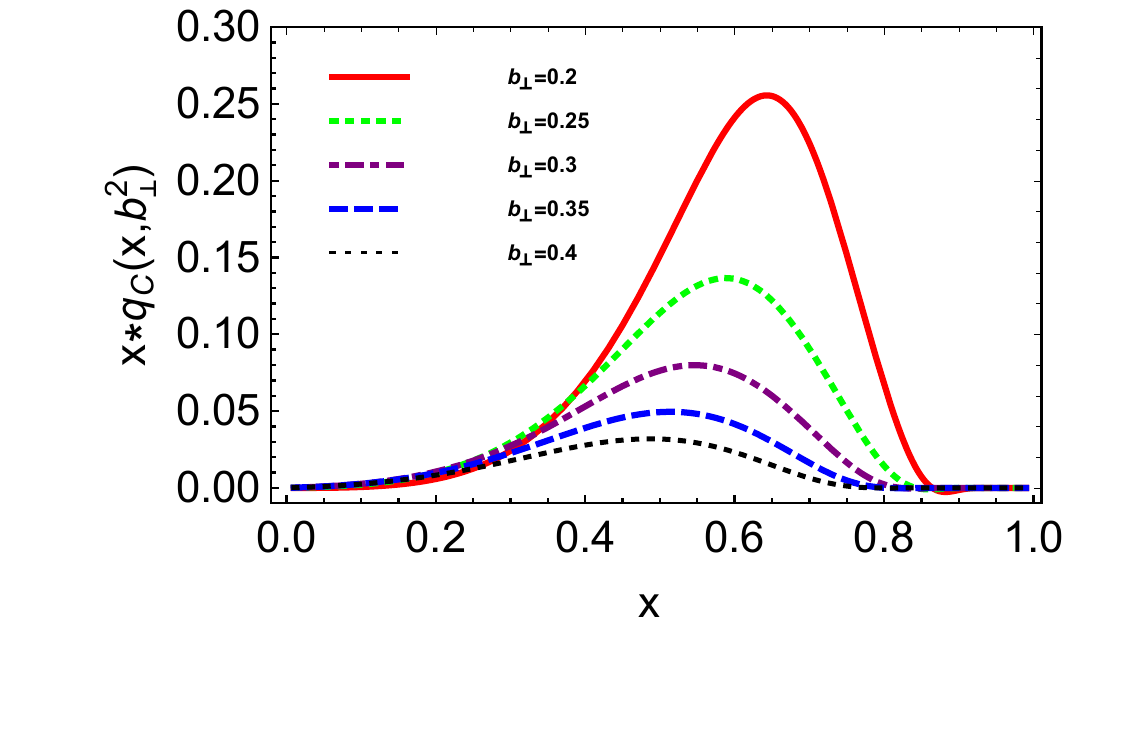}
\qquad
\includegraphics[width=0.47\textwidth]{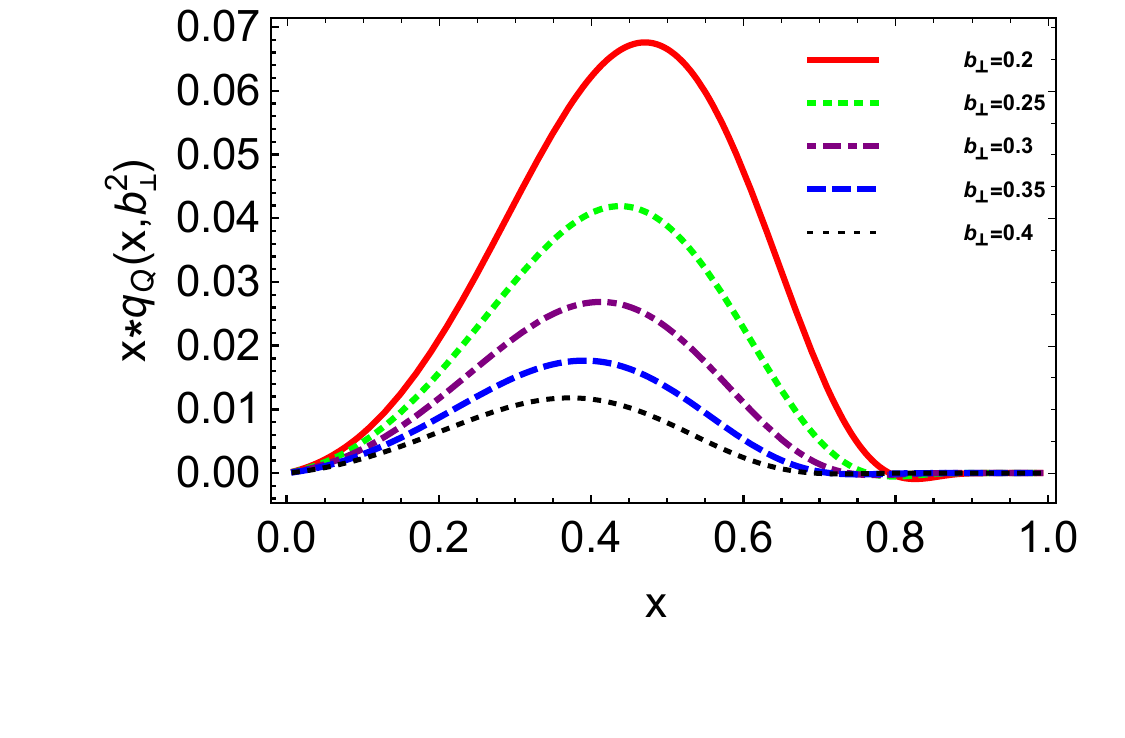}
\caption{The three distinct impact parameter-dependent PDFs are multiplied by the momentum fraction \( x \) for varying values of \( b_{\perp} \). The thick red solid line represents \( b_{\perp}=0.2 \) fm; the thick green dotted line corresponds to \( b_{\perp}=0.25 \) fm; the thick purple dot-dashed line indicates \( b_{\perp}=0.3 \) fm; the thick blue dashed line signifies \( b_{\perp}=0.35 \) fm; and the thin black dotted line illustrates \( b_{\perp}=0.4 \) fm.}\label{qmcq}
\end{figure}

The diagrams of \( x\cdot q_C \), \( x\cdot q_M \), and \( x\cdot q_Q \) are presented in Fig. \ref{qmcq}. The diagrams indicate that as \( b_{\perp} \) increases, the values of \( x\cdot q_{C,M,Q} \) decrease, while the corresponding \( x \)-value at which the peak occurs gradually diminishes. It is important to note that in the large \( x \) region, for small \( b_\perp \), the value of \( x \cdot q_Q \) exhibits oscillatory behavior around zero. The sign issue is of numerical origin. 



%
%
%

\subsection{The width distribution}

The distribution of parton widths for a specified momentum fraction $x$ is
\begin{align}\label{tmmf}
\langle \bm{b}_{\bot}^2\rangle_x &=\frac{\int \mathrm{d}^2 \bm{b}_{\bot}\bm{b}_{\bot}^2q(x,\bm{b}_{\bot}^2)}{\int \mathrm{d}^2 \bm{b}_{\bot}q(x,\bm{b}_{\bot}^2)},
\end{align}
when $x\rightarrow 1$, the impact parameter should approach zero. This is because the struck quark moves closer to the center of momentum as its momentum increases.
\begin{figure}
\centering
\includegraphics[width=0.47\textwidth]{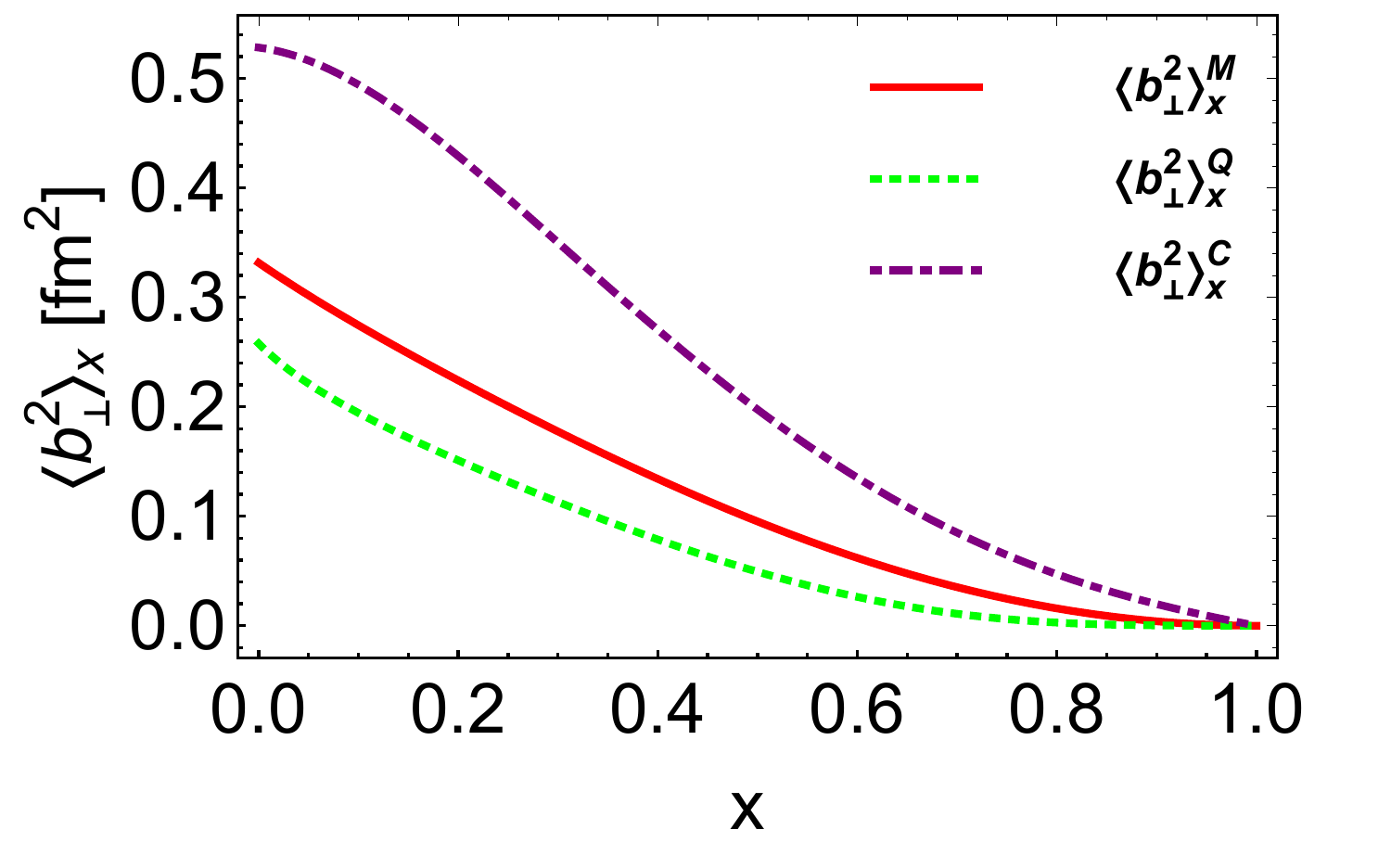}
\caption{The width distribution for a given momentum fraction \( x \) is defined in Eq. (\ref{tmmf}). The red solid line represents \( \langle \bm{b}_{\bot}^2\rangle_x^M \), the green dotted line denotes \( \langle \bm{b}_{\bot}^2\rangle_x^Q \), and the purple dot-dashed line illustrates \( \langle \bm{b}_{\bot}^2\rangle_x^C \).}\label{witf}
\end{figure}
\begin{figure}
	\centering
	\includegraphics[width=0.47\textwidth]{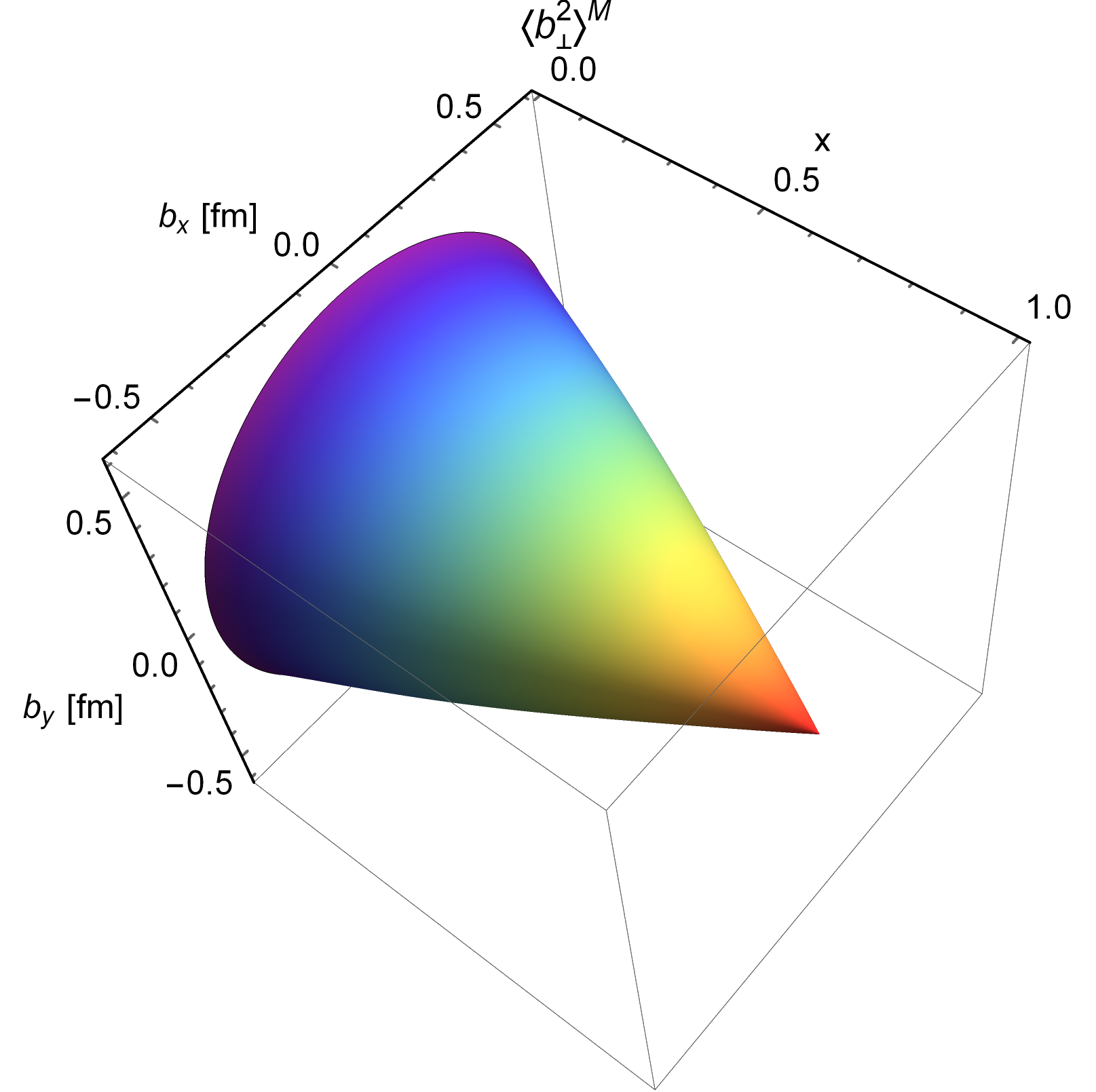}
	\qquad
	\includegraphics[width=0.47\textwidth]{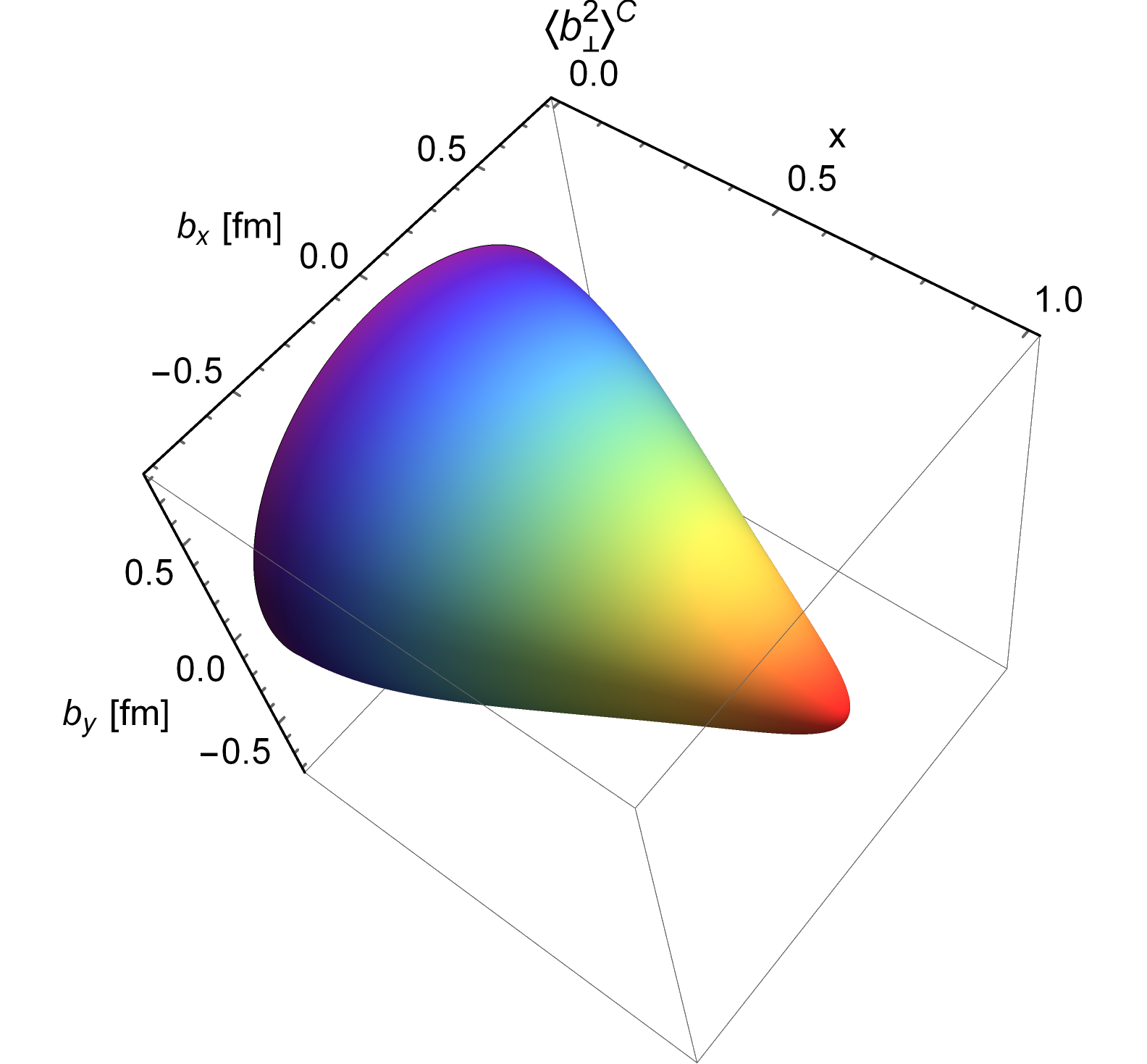}
	\qquad
	\includegraphics[width=0.47\textwidth]{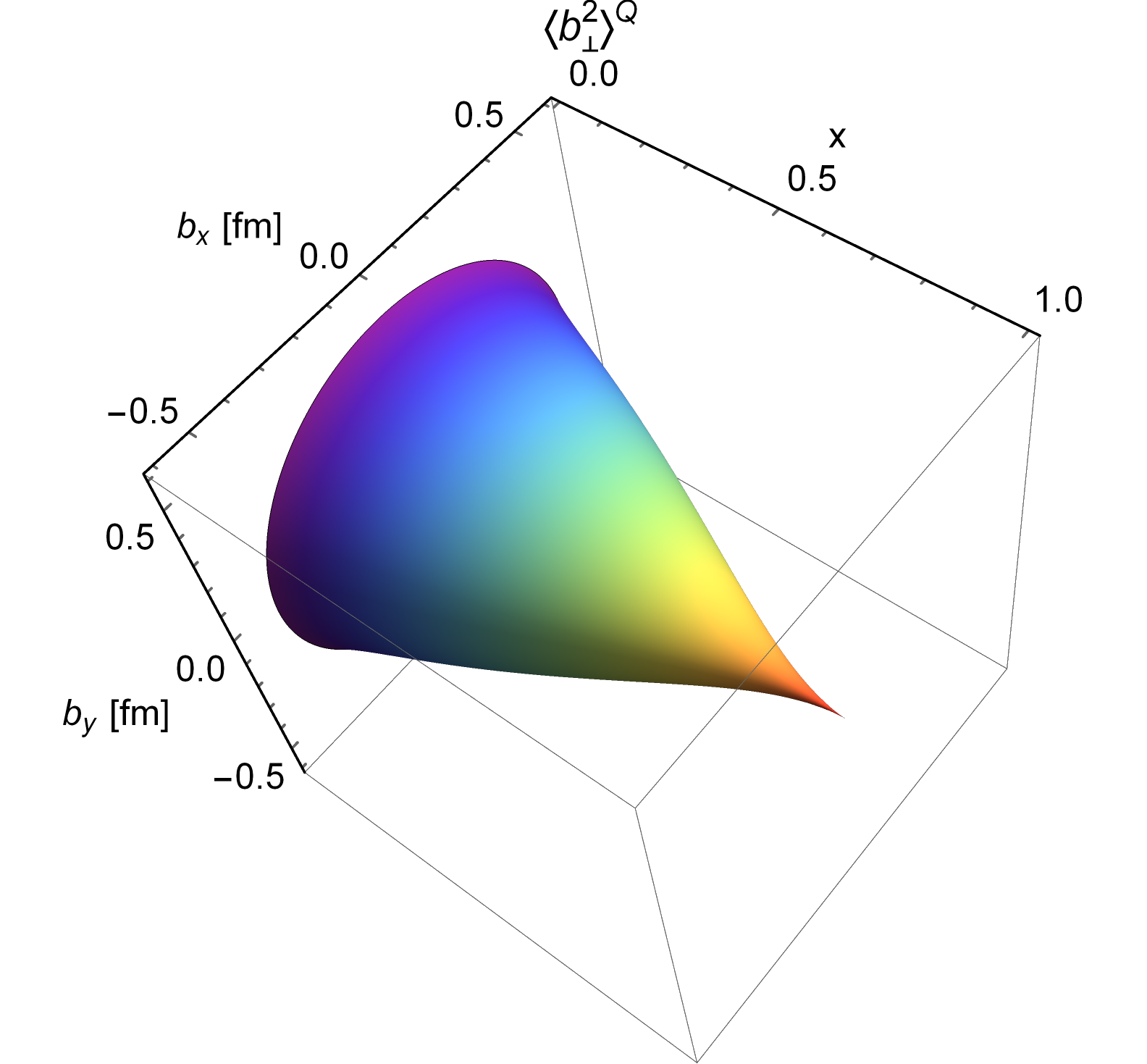}
	\caption{Three-dimensional representation of the function of Eq. (\ref{tmmf}), showing the $x$-dependence of the $\rho$ meson’s transverse magnetic, charge, and quadrupole radius.}\label{3dqcm}
\end{figure}

The width distributions for the three datasets are illustrated in Fig. \ref{witf}. It is evident that all of them satisfy the condition $\langle \bm{b}_{\bot}^2\rangle_1=0$. When $x$ is small, $\langle \bm{b}_{\bot}^2\rangle_x^C$ exhibits the largest value, while $\langle \bm{b}_{\bot}^2\rangle_x^Q$ shows the smallest. This indicates that in impact parameter space, the charge distribution is the broadest, whereas the quadrupole distribution is the narrowest.


In Fig. \ref{3dqcm}, we present a three-dimensional representation of the function described in Eq. (\ref{tmmf}). This illustration highlights the $x$-dependence of the transverse magnetic, charge, and quadrupole radii of the $\rho$ meson, as referenced in Ref.~\cite{Dupre:2017hfs}. The diagram reveals that the transverse quadrupole radius is minimized across the entire region where $x\in[0,1]$. Additionally, it is observed that the transverse charge radius exhibits its maximum breadth.  

The squared radius is determined from $\langle \bm{b}_{\bot}^2\rangle_x$ by computing the following average over $x$:
\begin{align}\label{tmmf1}
\langle \bm{b}_{\bot}^2\rangle &=\frac{\int_0^1 \mathrm{d}x\int \mathrm{d}^2 \bm{b}_{\bot}\bm{b}_{\bot}^2q(x,\bm{b}_{\bot}^2)}{\int_0^1 \mathrm{d}x\int \mathrm{d}^2 \bm{b}_{\bot}q(x,\bm{b}_{\bot}^2)}.
\end{align}
In Table \ref{tb41}, we present the $\langle \bm{b}_{\bot}^2\rangle_x$ at $x=0$, which indicates the ranges of various distributions. In impact parameter space, the range of charge distribution $q_C$ is the most extensive, while the quadrupole distribution $q_Q$ exhibits the narrowest range. The values of $\langle \bm{b}_{\bot}^2\rangle_0$ is the same magnitude as results of pseudoscalar meson in Refs.~\cite{Zhang:2021tnr,Zhang:2024adr} and quarks in the proton of Ref.~\cite{Dupre:2017hfs}

\begin{center}
\begin{table}
\caption{Comparison of the $x$-averaged squared radius for $\langle \bm{b}_{\bot}^2\rangle$ and the value of $\langle \bm{b}_{\bot}^2\rangle_x$ at $x=0$ in the NJL model in units of fm$^2$.}\label{tb41}
\begin{tabular}{p{1.2cm}p{1.1cm} p{1.1cm} p{1.1cm} p{1.1cm}p{1.1cm}p{1.1cm}}
\hline\hline
 &$\langle \bm{b}_{\bot}^2\rangle_0^{C}$&$\langle \bm{b}_{\bot}^2\rangle_0^{M}$&$\langle \bm{b}_{\bot}^2\rangle_0^{Q}$&$\langle \bm{b}_{\bot}^2\rangle^{C}$&$\langle \bm{b}_{\bot}^2\rangle^{M}$&$\langle \bm{b}_{\bot}^2\rangle^{Q}$\\
\hline
NJL &0.528&0.332&0.258&0.220&0.126&0.094\\
\hline\hline
\end{tabular}
\end{table}
\end{center}

\section{Summary and conclusion}\label{excellent}
In this study, we assess the form factors and impact parameter space parton distribution functions derived from the generalized parton distributions (GPDs) of the $\rho$ meson within the framework of the Nambu–Jona-Lasinio (NJL) model, employing proper time regularization. We select the NJL model due to its ability to provide qualitatively sound initial conditions for exploring physical possibilities in domains where more realistic frameworks have yet to yield insights or predictions.

For the Sachs-like charge, magnetic, and quadrupole form factors of the $\rho$ meson, we compare our results with lattice QCD data. The findings indicate that the dressed $G_C^D$ and $G_M^D$ align well with the lattice QCD data; however, for the quadrupole form factors, both the dressed $G_Q^D$ and bare $G_Q$ exhibit harder behaviors than those observed in lattice results.

We also examine the structure functions $A(Q^2)$, $B(Q^2)$, and tensor polarization $T_{20}(Q^2,\theta)$. The values obtained are in good agreement with those derived under various limits of $Q^2$. Furthermore, we investigate helicity-conserving matrix elements such as $G_{00}^+$ and $G_{11}^+$ alongside helicity non-conserving matrix elements like $G_{0+}^+$ and $G_{-+}^+$. Both bare and dressed cases have been studied, yielding results consistent with other findings.

The impact parameter dependent parton distribution functions (PDFs) are analyzed. The diagrams of \( x\cdot q_C \), \( x\cdot q_M \), and \( x\cdot q_Q \) are illustrated in Fig. \ref{qmcq}. A closer examination reveals several intriguing characteristics regarding the distributions of valence constituents within the \( \rho \) meson. The diagrams indicate that as \( b_{\perp} \) increases, the values of \( x\cdot q_{C,M,Q} \) exhibit a decreasing trend, while the corresponding \( x \)-value at which the peak occurs gradually diminishes.

We also investigate the width distributions corresponding to the three PDFs in impact parameter space. The radius of the magnetic distribution $q_M$ falls in between, situated between the charge distribution $q_C$ and quadrupole distribution $q_Q$.

To obtain a more realistic form factor for the $\rho$ meson, it is essential to consider the contributions from gluons. We can compute the Sachs-like charge, magnetic, and quadrupole form factors of the $\rho$ meson using Dyson-Schwinger equations (DSEs) and subsequently compare these results with those obtained from this model as well as lattice QCD data.

Additionally, we can evaluate the gravitational form factors, which related to higher Mellin moments of $\rho$ meson GPDs within the NJL model and verify whether these GPDs satisfy polynomial conditions. Furthermore, we will assess the twist-2 chiral odd quark transversity GPDs for the $\rho$ meson in the NJL model. Finally, we aim to examine both unpolarized,  polarized and transversity GPD polynomial sum rules associated with the $\rho$ meson.



\acknowledgments
Work supported by: the Scientific Research Foundation of Nanjing Institute of Technology (Grant No. YKJ202352), the Natural Science Foundation of Jiangsu Province (Grant No. BK20191472), and the China Postdoctoral Science Foundation (Grant No. 2022M721564).

\appendix
\section{Appendix 1: useful formulas}\label{AppendixT1}
Here we use the gamma-functions ($n\in \mathbb{Z}$, $n\geq 0$)
\begin{subequations}\label{cfun}
\begin{align}
\mathcal{C}_0(z)&:=\int_0^{\infty} \mathrm{d}s\, s \int_{\tau_{uv}^2}^{\tau_{ir}^2} \mathrm{d}\tau \, e^{-\tau (s+z)}\nonumber\\
&=z[\Gamma (-1,z\tau_{uv}^2 )-\Gamma (-1,z\tau_{ir}^2 )]\,, \\
\mathcal{C}_n(z)&:=(-)^n\frac{z^n}{n!}\frac{\mathrm{d}^n}{\mathrm{d}\sigma^n}\mathcal{C}_0(z)\,, \\
\bar{\mathcal{C}}_i(z)&:=\frac{1}{z}\mathcal{C}_i(z),
\end{align}
\end{subequations}
where $\tau_{uv,ir}=1/\Lambda_{\text{UV},\text{IR}}$ are, respectively, the infrared and ultraviolet regulators described above, with $\Gamma (\alpha,y )$ being the incomplete gamma-function.

%
%

\bibliographystyle{apsrev4-1}
\bibliography{zhangrho}

\end{document}